\documentclass[twoside,journey]{IEEEtran}
\usepackage{makecell}
\usepackage{array}
\usepackage{graphicx,amssymb,amsmath}
\usepackage{multicol}
\usepackage[noadjust]{cite}
\usepackage{setspace}
\usepackage{subfigure}
\usepackage{graphicx}
\usepackage{float}
\usepackage {url}
\usepackage{stfloats}
\usepackage{amsthm,pifont}
\usepackage{flushend}
\usepackage{cases,subeqnarray}
\usepackage{bm,multirow,bigstrut}
\usepackage{amsmath, amsthm, amssymb}
\usepackage{textcomp}
\usepackage{latexsym,bm}
\usepackage{booktabs}
\usepackage{xcolor}
\usepackage{mathtools}
\usepackage{dsfont}
\usepackage{extarrows}
\usepackage{epsfig}
\usepackage{epsfig}
\usepackage{epstopdf}
\usepackage[noend]{algpseudocode}
\usepackage{algorithmicx,algorithm}
\theoremstyle{plain}

\theoremstyle{plain}
\newtheorem{rem}{Remark}
\newtheorem{them}{Theorem}

\newtheorem{prop}{Proposition}
\usepackage{amsmath}
\newtheorem{theorem}{Theorem}
\newtheorem{lemma}{Lemma}
\IEEEoverridecommandlockouts
\begin{document}
	%----------------------------title&author&thanks----------------------------
	\title{Performance and Optimization of Reconfigurable Intelligent Surface Aided THz Communications}
	\author{Hongyang~Du, Jiayi~Zhang, Ke~Guan, Dusit~Niyato, Huiying Jiao, Zhiqin Wang, and Thomas K{\"u}rner
	\thanks{H.~Du is with the School of Computer Science and Engineering, the Energy Research Institute @ NTU, Interdisciplinary Graduate Program, Nanyang Technological University, Singapore (e-mail: hongyang001@e.ntu.edu.sg).}
	\thanks{J.~Zhang is with the School of Electronic and Information Engineering, Beijing Jiaotong University, Beijing 100044, China. (e-mail: jiayizhang@bjtu.edu.cn).}
	\thanks{K. Guan is with the State Key Laboratory of Rail Traffic Control and Safety, Beijing Jiaotong University, Beijing 100044, China. (e-mail: kguan@bjtu.edu.cn).}
	%\thanks{H. Xiao is with State Key Laboratory of Mobile Network and Mobile Multimedia Technology, China.}
	\thanks{D. Niyato is with the School of Computer Science and Engineering, Nanyang Technological University, Singapore (e-mail: dniyato@ntu.edu.sg)}
	\thanks{H. Jiao and Z. Wang are with China Academy of Information and Communications Technology, Beijing 100191, P. R. China.}
	\thanks{T. K{\"u}rner is with Institute for Communications Technology (IfN), Technische Universit{\"a}t Braunschweig, 38106 Brunswick, Germany. (e-mail: kuerner@ifn.ing.tu-bs.de)}
	}
	
	\maketitle
	%----------------------------abstract----------------------------
	\vspace{-1.7cm}
	\begin{abstract}
	TeraHertz (THz) communications can satisfy the high data rate demand with massive bandwidth. However, severe path attenuation and hardware imperfection greatly alleviate its performance. Therefore, we utilize the reconfigurable intelligent surface (RIS) technology and investigate the RIS-aided THz communications. We first prove that the small-scale amplitude fading of THz signals can be accurately modeled by the fluctuating two-ray distribution based on two THz signal measurement experiments conducted in a variety of different scenarios. To optimize the phase-shifts at the RIS elements, we propose a novel swarm intelligence-based method that does not require full channel estimation. We then derive exact statistical characterizations of end-to-end signal-to-noise plus distortion ratio (SNDR) and signal-to-noise ratio (SNR). Moreover, we present asymptotic analysis to obtain more insights when the SNDR or the number of RIS's elements is high. Finally, we derive analytical expressions for the outage probability and ergodic capacity. The tight upper bounds of ergodic capacity for both ideal and non-ideal radio frequency chains are obtained. It is interesting to find that increasing the number of RIS's elements can significantly improve the THz communications system performance. For example, the ergodic capacity can increase up to $25\%$ when the number of elements increases from $40$ to $80$, which incurs only insignificant costs to the system.
	\end{abstract}
	%----------------------------keywords----------------------------
	\vspace{-0.2cm}
	%\newpage
	\IEEEpeerreviewmaketitle
	%----------------------------introduction----------------------------
	\section{Introduction}
	In the past few years, wireless data traffic has been witnessing unprecedented expansion. Several technological advances, such as artificial intelligence and massive multiple-input multiple-output (MIMO) systems, have been presented \cite{agiwal2016next} while the wireless world moves towards the sixth generation (6G). Developing new spectrum resources for future 6G communications systems is urgently demanded.
	
	Because of the high spectrum availability, TeraHertz (THz) communications emerge as a key candidate in future generations \cite{boulogeorgos2018terahertz,chaccour2020risk,faisal2020ultramassive,chaccour2021seven}. However, THz signals have considerably high sensitivity to blockage, i.e., in urban environments with high obstacle density. Thus, the direct usage of THz-band in wireless access scenarios is challenging. Recently, the reconfigurable intelligent surface (RIS) technology has drawn increasing attention \cite{basar2019wireless,wu2019intelligentsda} due to its ability to improve both spectrum and energy efficiency. RIS is based on massive integration of low-cost tunable passive elements that can be easily coated on existing infrastructures such as walls of buildings, facilitating low-cost and low-complexity implementation \cite{liu2020energy,alegria2019achievable,boulogeorgos2020how,xing2020achievable}. By smartly adjusting the scattering properties of its reflecting elements, the RIS can form equally strong beams in any direction for THz signals, which leads to a more controllable wireless environment. Compared to conventional relays, RISs can enhance the transmission performance without being impaired by self-interference and thermal noise. More importantly, RIS can be deployed to leverage the line-of-sight (LoS) components with respect to both the base station (BS) and the users to assist their end-to-end THz communications, which potentially reduce the path attenuation that originates from the high frequencies.
	
	However, how to model the small-scale amplitude fading of THz signals in the RIS-aided system is still an open question. Methods of modeling and evaluating the performance of THz communications have been extensively studied in the literature \cite{tekbiyik2020reconfigurable,maruthi2020improving,boulogeorgos2021cascaded,ma2020joint,qiao2020secure,boulogeorgos2020performance,jung2019reliability,kokkoniemi2018simplified,boulogeorgos2018performance}. For example, the authors in \cite{kokkoniemi2018simplified} delivered a simplified molecular absorption loss model that was used in \cite{boulogeorgos2018performance} for the quantification of the THz link capacity. However, the THz system performance is overestimated by neglecting small-scale fading. Although the small-scale fading is taken into consideration in \cite{boulogeorgos2020performance,jung2019reliability}, the channel modelings were not based on real THz signal measurements. Furthermore, in \cite{ekti2017statistical}, the envelope of the fading coefficient was proved by the authors experimentally to follow the Nakagami-$m$ distribution under both non-line-of-sight (NLoS) and LoS conditions. Similarly, experimental verification of the existence of shadowing in the $300$ GHz band were exploited in \cite{priebe2011channel}. Very recently, a THz measurement campaign was conducted in a train test center with trains, tracks, and lampposts \cite{KeT2TTrans}. We fit the measurement data to Gaussian, Rician, Nakagami-$m$, and fluctuating two-ray (FTR) distributions \cite{romero2017fluctuating}, respectively, and prove that the FTR distribution, which is a generalization of the two-wave with diffuse power fading model and allows the constant amplitude specular waves of LoS propagation to fluctuate randomly, performs a much better fit than others.
	
	Therefore, with the help of the FTR fading model, we investigate the RIS-aided THz communications. We propose an algorithm to obtain the optimal design of RIS's reflector array. By designing the phase shift, we can then theoretically derive the maximum of SINR and corresponding probability density function (PDF) and cumulative distribution function (CDF) expressions. The main contributions of this paper are summarized as follows:
	\begin{itemize}
	\item To model the small-scale amplitude fading of THz signals, we prove that the FTR distribution performs a much better fit than those of the other fading models with the help of the THz measurement data \cite{KeT2TTrans,papasotiriou2021experimentally}. Moreover, we derive exact PDF and CDF expressions of end-to-end SNR and SNDR in terms of the multivariate Fox's $H$-function, respectively. The impact of transceiver antenna gains, the level of transceiver hardware imperfections, the transmission distance, the misalignment between the RIS and the destination, the operating frequency, and the environmental conditions including temperature, humidity, and pressure, are all comprehensively considered in our analysis.
	\item For the practical RIS with a finite number of phase-shifters at each element, we propose a swarm intelligence-based optimization (SIO) algorithm to obtain the optimal phase-shifts. Furthermore, the impact of the number of discrete phase-shift levels on the transmission performance is revealed, and we find that RIS with discrete phase-shifts can achieve performance close to that of the continuous phase-shifts.
	\item We derive exact closed-form expressions of outage probability for the RIS-aided THz communications system with both ideal and non-ideal radio frequency (RF) chains. Moreover, asymptotic analysis is presented when the SNDR or the number of RIS's elements is high, which is easier to calculate and obtain insights. For the ergodic capacity, the exact closed-form expression for ideal RF chains is obtained and the tight upper bound expressions for both ideal and non-ideal RF chains are derived. Important physical and engineering insights have been revealed. The derived results show that the propagation loss is more detrimental than the molecular absorption loss, and the performance is improved significantly by increasing the number of reflecting elements equipped in an RIS.
	\end{itemize}
	
	The remainder of the paper is structured as follows. We briefly introduce the RIS-aided THz communications system and signal models in Section \ref{sysmodel}. Section \ref{afefaegva} presents the optimal phase-shifts of the RIS's reflector array under discrete level constraints. In Section \ref{eagf34}, we derive exact PDF and CDF expressions of end-to-end SNR and SNDR of the considered system. Section \ref{casfae341} presents the asymptotic analysis to obtain several insights. Important performance metrics, such as outage probability and ergodic capacity, are obtained in Section \ref{casfae34}. Numerical results accompanied with Monte-Carlo simulations are presented in Section \ref{nad}. Finally, the conclusions are drawn in Section \ref{cons}.
	%----------------------------system model----------------------------
	\section{System and Signal Models}\label{sysmodel}
	\subsection{System Description}
	\begin{figure}[t]
	\centering
	\includegraphics[width=.45\textwidth]{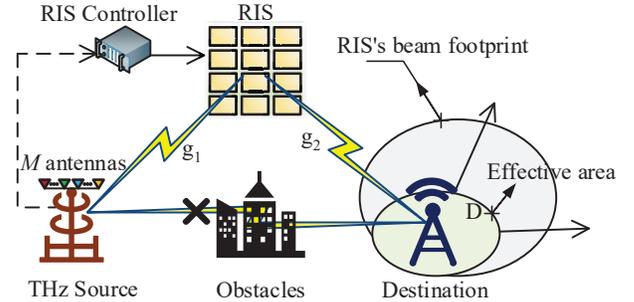}
	\caption{System model of the RIS-aided THz communications.}
	\label{model}
	\end{figure}
	We focus on a single cell of the THz communications system. As shown in Fig. \ref{model}, the Source is equipped with $M$ antennas, and the Destination has one antenna.  The interference among THz links can be significantly small because the signal in THz communications is highly directional. Thus, the interference from other cell-users is ignored. Because THz signals suffer from high path and penetration losses, an RIS is set up to assist the communications. The RIS is a uniform planar array (UPA) containing $L$ nearly passive reconfigurable reflecting units, where each element is capable of electronically controlling the phase of the incident electromagnetic waves. To establish a practical RIS-aided THz communications model, we consider the following coefficients and give the corresponding reasons:	
	\begin{itemize}
	\item {\it Small-Scale Fading Coefficient ${\mathbf h}_F$}: Although some of the existing literature on RIS-aided THz communications considers the utilization of small-scale fading models, classical models such as Rayleigh and Rician fadings are typically used. However, recent real measurements of THz signals show that these classical models cannot fit well the THz signals. Although the authors of \cite{{papasotiriou2021experimentally}} claim that $\alpha\!-\!\mu$ fading obtains good results and passes the Kolmogorov-Smirnov (K-S) test, we can observe that the $\alpha\!-\!\mu$ fading is still not accurate enough in fitting the bimodal characteristics of THz signal's amplitude fading. Therefore, a more accurate channel model is needed.
	\item {\it Misalignment Fading Coefficient $h_P$:} Random building sways, which will be caused by wind loads and thermal expansions, will result in pointing errors at Destination. Because the THz signals are highly directional, we need to take the misalignment into consideration.
	\item {\it Path Gain Coefficient $h_L$:} Because of the THz signals' inability to penetrate solid objects, the path gain in the THz communications system concludes the propagation gain and the molecular absorption gain.
	\item {\it Hardware Impairments $n_S$ and $n_D$:} Physical transceivers have hardware impairments that create distortions that degrade the performance of communications systems. The vast majority of technical contributions in the area of relaying neglect hardware impairments and, thus, assume ideal hardware. Such approximations make sense in low-rate systems, but can lead to very misleading results when analyzing future high-rate systems. This paper quantifies the impact of hardware impairments on the RIS-aided THz system.
	\end{itemize}
	By considering the above coefficients and the characteristics of RIS carefully, we can derive the expression of the received signal, which allows us to further analyze the statistical characteristics of SNR and performance metrics. A more detailed analysis of the above coefficients are given as follows:
	
	\subsubsection{Small-Scale Fading Measurement and Modeling}\label{fitt}
	We consider the RIS is electrically small and the correlation between the amplitudes of the signals is weak enough to be ignored \cite{atapattu2020reconfigurable,peng2021multiuser}. Let $ {{\mathbf{g}}_1} \in {\mathbb{C}^{M \times L}} $ and $ {{\mathbf{g}}_2} \in {\mathbb{C}^{1 \times L}} $ denote the channels from the THz Source to the RIS, and from the RIS to the Destination, respectively, $ {\bf{\Phi }} = \beta {\rm{diag}}\left( {\bf{\varphi}}  \right)$ denote the diagonal reflection coefficients matrix of the RIS, with $ \beta  \in (0,1] $ is the phase-shifts matrix\footnote{The value of $\beta$ does not affect the algorithm design. Without loss of generality, we set $\beta=1$ for simplicity.}. The small-scale fading between BS and RIS is defined as $ {\mathbf{h}_F} \triangleq {{\mathbf{g}}_2}{\bf{{\Phi }}}{{\mathbf{g}}_1^{\rm H}} $, where {\small $ {\bf{\varphi}}  = {\left( {{e^{i{\varphi _1}}}, \ldots ,{e^{i{\varphi _L}}}} \right)^T} $}, {\small $\varphi _\ell$} is the phase-shift of the $\ell_{\rm th}$ element of the RIS, {\small $i={\sqrt{-1}}$}, {\small $ {{\bf{g}}_2} = \left( {g_2^{\left( 1 \right)}, \ldots ,g_2^{\left( L \right)}} \right) = \left( {\left| {g_2^{\left( 1 \right)}} \right|{e^{ - i\theta _2^{\left( 1 \right)}}}, \ldots ,\left| {g_2^{\left( L \right)}} \right|{e^{ - i\theta _2^{\left( L \right)}}}} \right) $}, and
	{\small \begin{align}
	{{\bf{g}}_1}& = \left( {\begin{array}{*{20}{c}}
	{g_1^{\left( {1,1} \right)}}& \cdots &{g_1^{\left( {1,L} \right)}}\\
	\vdots & \ddots & \vdots \\
	{g_1^{\left( {M,1} \right)}}& \cdots &{g_1^{\left( {M,L} \right)}}
	\end{array}} \right)
\notag\\
&= \left( {\begin{array}{*{20}{c}}
	{\left| {g_1^{\left( {1,1} \right)}} \right|{e^{ - i\theta _1^{\left( {1,1} \right)}}}}& \cdots &{\left| {g_1^{\left( {1,L} \right)}} \right|{e^{ - i\theta _1^{\left( {1,L} \right)}}}}\\
	\vdots & \ddots & \vdots \\
	{\left| {g_1^{\left( {M,1} \right)}} \right|{e^{ - i\theta _1^{\left( {M,1} \right)}}}}& \cdots &{\left| {g_1^{\left( {M,L} \right)}} \right|{e^{ - i\theta _1^{\left( {M,L} \right)}}}}
	\end{array}} \right).
	\end{align}}\noindent
	The elements in ${{\mathbf{g}}_1}$ and ${{\mathbf{g}}_2}$ represent the small-scale fading of the THz signals, for which we have to find the appropriate channel fading model. Recently, the FTR fading \cite{romero2017fluctuating} has received a great amount of research interest, because the FTR model yields a better fit than those of the other conventional channel models with experimental data in high-frequency communications. The PDF of the squared FTR RV $\gamma$ can be expressed as \cite{zhang2017new}
	{\small \begin{equation}\label{PDFFTR}
	{f_\gamma }\left( \gamma \right) = \frac{{{m^m}}}{{\Gamma (m)}}\sum\limits_{j = 0}^{N_T}  {\frac{{{K^j}{d_j}}}{{j!}}}\frac{{{\gamma^j}}}{{\Gamma (j + 1){{\left( {2{\sigma ^2}} \right)}^{j + 1}}}}\exp \left( { - \frac{\gamma}{{2{\sigma ^2}}}} \right),
	\end{equation}	}\noindent
	where
	{\small \begin{align}
	&{d_n} \buildrel \Delta \over = \sum\limits_{k = 0}^n {\left(\! {\begin{array}{*{20}{l}}
	n\\
	k
	\end{array}} \!\right)} {\left(\! {\frac{\Delta }{2}} \!\right)^k}\sum\limits_{l = 0}^k {\left( {\begin{array}{*{20}{c}}
	k\\
	l
	\end{array}} \right)} \Gamma\!(n + m + 2l - k){e^{\frac{{\pi (2l - k)j}}{2}}}\notag\\
	&\times\!\!{\left(\! {{{(m\! +\! K)}^2}\! -\! {{(K\Delta )}^2}} \!\right)^{\frac{{ - (n + m)}}{2}}}\!P_{n + m - 1}^{k - 2l}\left(\!\! {\frac{{m + K}}{{\sqrt {{{(m\! + \!K)}^2} \!- \!{{(K\Delta )}^2}} }}} \!\!\right)\!,
	\end{align}}\noindent
	where $P_ \cdot ^ \cdot \left(  \cdot  \right)$ denotes the Legendre function of the first kind \cite[eq. (8.702)]{gradshteyn2007}, $m$ is the fading severity parameter, $K$ is the average power ratio of the dominant wave to the scattering multipath, and $\Delta $ is a parameter varying from $0$ to $1$ representing the similarity of two dominant waves. The received average SNR is given by $ \upsilon =2 \sigma^{2}(1+K)$. Note that $N_T$ is the truncation term. Using the property of PDF, we have $ \int_0^\infty  {{f_\gamma }\left( \gamma  \right)}  = 1 $. With the help of \cite[eq. (3.351.3)]{gradshteyn2007} and \cite[eq. (8.339.1)]{gradshteyn2007}, the $N_T$ should be set to satisfy $ \sum\limits_{j = 0}^{{N_T}} {{d_j}}  \to 1 $. Typically, we can use $40$ terms to make the series converge for all cases considered in this paper. Furthermore, the truncation error, which is defined as $ 1-\sum\limits_{j = 0}^{{N_T}} {{d_j}}$, is less than $10^{-6}$.
	
	{To prove that the FTR model can fit the THz signal's amplitude fading best, we consider the real measurement data from two campaigns \cite{KeT2TTrans,papasotiriou2021experimentally}, and evaluate the fitting of those theoretical fading distributions to the empirical data with the help of K-S goodness-of-fit statistical test \cite{papoulis2002probability}.
	\begin{itemize}
	\item Very recently, a measurement campaign was conducted in a train test center \cite{KeT2TTrans}. The authors use a novel $M$-sequence correlation-based ultra-wideband channel sounder to measure the THz channel. In the measurement, the channel sounder was operated at $304.2$ GHz with $8$ GHz bandwidth. The small-scale fading was extracted for each channel impulse response based on the data. We have fitted the measurement data to a Gaussian, Rician, Nakagami-$m$, and FTR distribution, respectively, which are also compared with the CDF of the measurement data. As shown in Figs. \ref{PreviousFit}, we find that the FTR distribution performs the best fit. 
	\item In \cite{papasotiriou2021experimentally}, three sets of experimental measurements, which have been conducted in a shopping mall, an airport check-in area, and an entrance hall of a university towards different time periods, are used to accurately model the THz signal fading distribution. Analysis shows that conventional distributions, e.g., Rayleigh, Rice, and Nakagami-$m$, lack fitting accuracy, whereas the $\alpha\!-\!\mu$ distribution has an almost-excellent fit. However, we can observe that the $\alpha\!-\!\mu$ distribution is still deficient to show the characterizations of the experimental results: 1) The experimental PDF appears bimodality in many test environments, e.g., Transmitter 4 (${\rm{TX}}_{4}$) in the entrance hall or the shopping mall (see Fig. 2 in \cite{papasotiriou2021experimentally}), which cannot be fitted by the PDF of $\alpha\!-\!\mu$ distribution. 2) As the variable $x$ increases, the slope of CDF gradually increases and then decreases to $0$, but the CDF of $\alpha\!-\!\mu$ distribution cannot fit the process of the increasing slope of CDF when x is small. As shown in Fig. \ref{NewFit}, we can observe that the FTR fading is better than the $\alpha\!-\!\mu$ fading. Furthermore, although both the $\alpha\!-\!\mu$ and FTR distribution can pass the K-S test, the values for the FTR distribution is always smaller.
	\end{itemize}
	\begin{figure}[t]
	\centering
	\includegraphics[width=.45\textwidth]{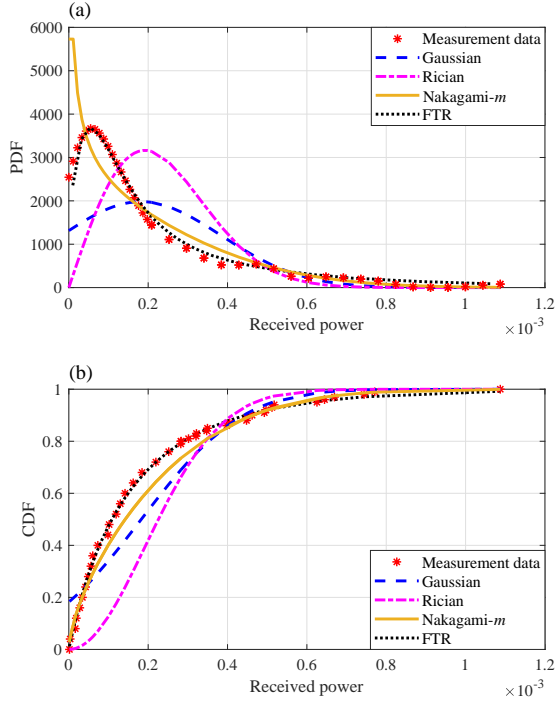}
	\caption{(a). PDF of the measurement data \cite{KeT2TTrans} and fitted distributions. The parameters of fitted distributions are obtained by ${\rm fitdist( \cdot )}$ function in Matlab \cite{web2} and are shown in Table \ref{tab1} (b). CDF of the measurement data and fitted distributions with the same parameters as Fig. \ref{PreviousFit} (a).}
	\label{PreviousFit}
	\end{figure}
	\begin{figure}[t]
	\centering
	\includegraphics[width=.45\textwidth]{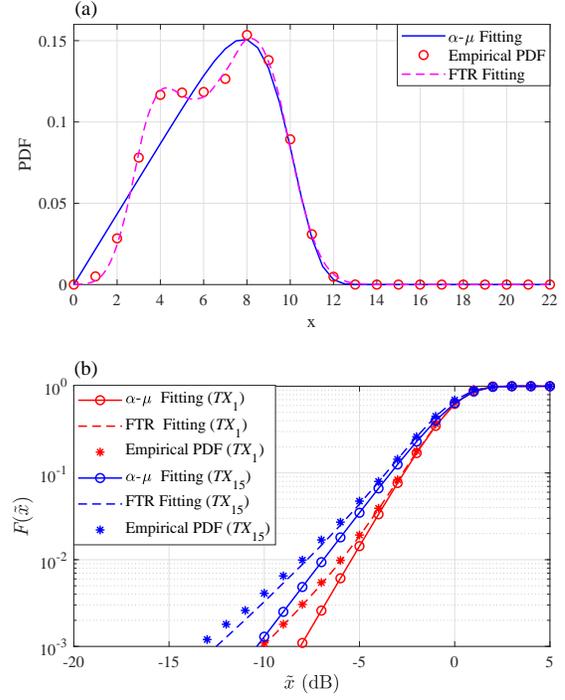}
	\caption{(a). PDF of the measurement data \cite{papasotiriou2021experimentally} and fitted distributions. The parameters of $\alpha\!-\!\mu$ distribution are given in \cite[Table 5, ${\rm{TX}}_{4}$]{papasotiriou2021experimentally}, i.e., $\alpha=8.54867$, $\mu=0.23377$, and $\beta=8.49874$. The parameters of FTR distribution are $ \sigma=0.8991$, $m=30$, $K=30$, and $\Delta =0.8$ (b). CDF of the measurement data \cite{papasotiriou2021experimentally} and fitted distributions. The parameters of $\alpha\!-\!\mu$ distribution are given in \cite[Table 1, ${\rm{TX}}_{1}$ and ${\rm{TX}}_{15}$]{papasotiriou2021experimentally}. The parameters of FTR distribution are shown in Table \ref{tab1}.}
	\label{NewFit}
	\end{figure}

	\begin{table}[h]
	\caption{{\label{tab1}Kolmogorov–Smirnov goodness of fit tests.}}
	\centering
	{\small \begin{tabular}{|p{1.4cm}|p{1.8cm}|p{2.37cm}|p{1.1cm}|}
	\toprule
	\hline
	Measurement   & Channel Model  & Parameters     & K-S Test Values \\ \hline
	\cite{KeT2TTrans}    & FTR Fading & $ \sigma=4.84\times10^{-3}$, $m=0.63$, $K=2.9$, $\Delta =0.3$  & $0.0383$ ($\sqrt{} $)  \\
	\hline
	\cite{KeT2TTrans}    & Gaussian Fading & $ {\mu _G}=1.83\times10^{-4}$, $ {\sigma _G}=2.01\times10^{-4}$   & $0.1870$ ($ \times $) \\
	\hline
	\cite{KeT2TTrans}      & Rician Fading &  $ {s_R}=7.78\times10^{-6}$, $ {\sigma _R}=1.92\times10^{-4}$    & $0.3687$ ($ \times $)  \\
	\hline
	\cite{KeT2TTrans}    & Nakagami-$m$ Fading &  ${\mu _N}=0.323$, ${\omega _N}=7.35\times10^{-8}$  & $0.1058$ ($ \times $)  \\
	\hline
	\cite{papasotiriou2021experimentally} ${\rm{TX}}_{1}$  & FTR Fading & $ \sigma=2.7729$, $m=6$, $K=7$, $\Delta =0.2$  & $0.0175$ ($\sqrt{}$) \\
	\hline
	\cite{papasotiriou2021experimentally} ${\rm{TX}}_{1}$  & $\alpha\!-\!\mu$ Fading & $\alpha=3.35467$, $\mu=1.11365$, and $\beta=11.58104$   & $0.0183$ ($\sqrt{}$)   \\
	\hline
	\cite{papasotiriou2021experimentally} ${\rm{TX}}_{15}$ & FTR Fading & $\sigma=2.382$, $m=3$, $K=6$, $\Delta =0.04$ & $0.0203$ ($\sqrt{}$) \\
	\hline
	\cite{papasotiriou2021experimentally} ${\rm{TX}}_{15}$ & $\alpha\!-\!\mu$ Fading  & $\alpha=3.11872$, $\mu=0.91962$, and $\beta=9.8334$   & $0.0566$ ($\sqrt{}$)\\
	\hline
	\bottomrule	
	\end{tabular}}
	\end{table}
	
	Moreover, Table \ref{tab1} shows that The FTR distribution achieves the smallest K-S test values in the fit of the experimental results of each THz signal measurement. Thus, in the following, we use the FTR fading model to illustrate the small-scale amplitude fading of THz communications.}
	
	\subsubsection{Misalignment Fading Coefficient}
	We consider that Destination has a circular detection beam, and the radius of the beam is $\alpha$. Furthermore, the RIS has a circular beam of radius $ \rho  $, where $ {\rm{0}} \le \rho  \le {\omega _d} $, and $w_{d}$ denotes the maximum radius of the beam. Moreover, we assume that $r$ is the pointing error. Due to the symmetry of the beam shapes, the misalignment fading coefficient that represents the fraction of the power collected by Destination, $h_P$, depends only on the radial distance $r$. Thus, $h_P$ can be approximated as \cite[eq. (11)]{farid2007outage}
	{\small \begin{align}\label{hppdfsAD}
	{f_{{h_P}}}(x) = {\gamma ^2}A_o^{ - {\gamma ^2}}{x^{{\gamma ^2} - 1}},\:\:\:\:\:0 \leqslant x \leqslant {A_o},
	\end{align}}\noindent
	where ${A_o} = {\left[ {{\mathop{\rm erf}\nolimits} \left(u \right)} \right]^2}$, $u\triangleq{\sqrt{\pi} a}/({\sqrt{2} w_{d}})$, $\gamma=\frac{w_{e q}}{2 \sigma_{S}}$, $\sigma_{S}^{2}$ denotes the variance of the pointing error displacement at Destination, ${\omega _{eq}}$ is the equivalent beam-width, and $w_{eq}^2 = w_d^2\sqrt \pi  \operatorname{erf} (u){\left( {2u\exp \left( { - {u^2}} \right)} \right)^{ - 1}}$.
	
	\subsubsection{Path Gain Coefficient}
	The path gain coefficient can be evaluated as ${h_L} = {h_{FL}}{h_{AL}}$, where ${h_{FL}}$ is the propagation gain, and ${h_{AL}}$ is the molecular absorption gain. Using Friis equation, we can express $h_{FL}$ as
	{\small \begin{equation}\label{hlsdwe}
	h_{FL}=\frac{c^2 \sqrt{G_{t} G_{r}}}{\left(4 \pi f\right)^2 \left( d_1d_2\right) },
	\end{equation}}\noindent
	where $G_{t}$ and $G_{r}$ represent the antenna orientation dependent transmission and reception gains, respectively, $c$ is the speed of light, $f$ stands for the operating frequency, $d_1$ is the distance between Source and RIS, and $d_2$ is the distance between RIS and Destination.
	Furthermore, ${h_{AL}}$ can be evaluated as \cite{kokkoniemi2018simplified}
	{\small \begin{equation}
	h_{AL}=\exp \left(-\frac{1}{2} \kappa_{\alpha}(f) \left(d_1+d_2\right)\right),
	\end{equation}}\noindent
	where $\kappa_{\alpha}(f)$ denotes the absorption coefficient which can be expressed as \cite{kokkoniemi2018simplified}	
	{\small \begin{align}
	{\kappa _\alpha }(f) = & \frac{{A(v)}}{{B(v) + {{\left( {\frac{f}{{100c}} - {c_1}} \right)}^2}}} + \frac{{C(v)}}{{D(v) + {{\left( {\frac{f}{{100c}} - {c_2}} \right)}^2}}} 
	\notag \\ &
	+ {p_1}{f^3} + {p_2}{f^2} + {p_3}f + {p_4},
	\end{align}}\noindent
	where $c$ denotes the speed of light, $c_{1}=10.835$ $\mathrm{cm}^{-1}$, $c_{2}=12.664$ $\mathrm{cm}^{-1}$, $p_{1}=5.54 \times 10^{-37}$ $\mathrm{Hz}^{-3}$, $p_{2}=-3.94 \times 10^{-25}$ $\mathrm{Hz}^{-2}$, $p_{3}=9.06 \times 10^{-14}$ $\mathrm{Hz}^{-1}$, $p_{4}=-6.36 \times 10^{-3}$, $ A(v) = 0.2205v(0.1303v + 0.0294) $, $ B(v) = {(0.4093v + 0.0925)^2} $, $ C(v) = 2.014v(0.1702v + 0.0303) $, and $ D(v) = {(0.537v + 0.0956)^2} $ \cite{kokkoniemi2018simplified}. Moreover, $v$ can be evaluated as	
	{\small \begin{equation}\label{afe}
	v \!=\! \phi \!\left( {0.06116{p^{ - 1}}\! +\! 2.1148 \!\times\! {{10}^{ - 7}}} \right)\exp\! \left(\! {\frac{{17.502T}}{{240.97 \!+\! T}}} \!\right),
	\end{equation}}\noindent
	where $T$ is the temperature, $\phi$ and $p$ denote the relative humidity and the atmospheric pressure, respectively. Note that the molecular absorption loss model that we considered here provides an accurate model for the $275$ – $400$ {\rm GHz} frequency band. Meanwhile, for frequencies higher than $400$ {\rm GHz}, we can obtain the value of $\kappa_{\alpha}(f)$ from the high resolution transmission (HITRAN) database \cite{chaccour2020can,babikov2012hitran}. Using \eqref{hlsdwe}-\eqref{afe}, we finally obtain the path gain coefficient of the considered RIS-aided THz communications system.

	\subsection{Received THz Signal Expression Analysis}
	The signal received from Source through RIS for Destination can be expressed as\footnote{Because the THz signals have high susceptibility to blockages \cite{chaccour2020can}, we assume that the direct link between the source and destination is negligible due to the severe attenuation.}
	{\small \begin{equation}\label{signal2}
	y =P {h_L}{h_P}{{\mathbf h}_F}{\bf{w}^T} \left( {x + {n_S}} \right) + {n_D} + n,
	\end{equation}}\noindent
where $n\sim {\mathcal{CN}}\left( 0,N_0\right)$, ${\mathcal{CN}}\left( {t_1,t_2} \right)$ denotes the white Gaussian process with mean $t_1$ and variance $t_2$, $x$ is the transmitted signal satisfying $ {\mathbb E}\left[ {{{\left| x \right|}^2}} \right] = 1 $, ${\bf{w}}\in {\mathbb{C}^{M \times 1}}$ is the beamforming vector, $ {\mathbb E}\left[ {{\left\| {{{\bf{w}}}} \right\|^2}} \right] = 1 $, $\left\| \cdot \right\|$ denotes the Frobenius norm, $P$ is the transmit power, the parameters ${n_S} \sim {\mathcal{CN}}\left( {0,\kappa _S^2P} \right)$ and ${n_D} \sim {\mathcal{CN}}\left( 0,\kappa_D^2P{{\left|{h_L}{h_P}{h_F} \right|}^2} \right)$ are distortion noise from hardware impairments in both Source and Destination, respectively \cite{bjornson2013newwqe}, and $ {\kappa _S} $ and $ {\kappa _D} $ are non-negative parameters that characterize the level of hardware impairments \cite{jensen2013robust} in Source and Destination, respectively. In the RIS-aided communications, the $P$ and ${\bf{\Phi}}$ are adjustable. Considering the optimal beamforming vector design, e.g., using the maximum ratio transmitting \cite{lo1999maximum,tokgoz2006performance}, we can define ${\bf{w}}$ as ${\bf{w}} =  {{{\mathbf{h}}_F}}/\left\| {{{\mathbf{h}}_F}} \right\|$, and the corresponding small-scale fading coefficient can be expressed as
	{\small \begin{align}
	{{{\mathbf{h}}_F}} &=\! \left(\! {{h_{F,1}},\!\ldots \!,{h_{F,M}}} \!\right) \!
	\notag\\&
	= \left(\! {\left| {\sum\limits_{\ell  = 1}^L {g_2^{\left( \ell  \right)}{e^{j{\varphi _\ell }}}g_1^{\left( {1,\ell } \right)}} } \right|, \!\ldots \!,\left| {\sum\limits_{\ell  = 1}^L {g_2^{\left( \ell  \right)}{e^{j{\varphi _\ell }}}g_1^{\left( {M,\ell } \right)}} } \right|} \!\right)\!\!.
	\end{align}}\noindent
	The square of the Frobenius norm of ${{{\mathbf{h}}_F}}$ is upper-bounded as
	{\small \begin{equation}\label{upperbou}
	{\left\| {{{\mathbf{h}}_F}} \right\|^2}\! = \!\!{\left(\! {\sum\limits_{m = 1}^M {\left| {\sum\limits_{\ell  = 1}^L {g_2^{\left( \ell  \right)}{e^{j{\varphi _\ell }}}g_1^{\left( {1,\ell } \right)}} } \right|} } \!\right)^2}\!\! \leqslant \!\!{\left(\! {\sum\limits_{m = 1}^M {\sum\limits_{\ell  = 1}^L {\left| {g_2^{\left( \ell  \right)}} \right|\left| {g_1^{\left( {m,\ell } \right)}} \right|} } } \!\right)^2}\!,
	\end{equation}}\noindent
	where the error between the upper-bound and the exact value of ${\left\| {{{\mathbf{h}}_F}} \right\|^2}$ decreases as RIS's phase shift design becomes more accurate. However, the right-hand side of \eqref{upperbou} is still hard to be used in the performance analysis. The reason is that $\left| {g_2^{\left( \ell  \right)}} \right|$ and $\left| {g_1^{\left( {m,\ell } \right)}} \right|$ follow i.n.i.d. FTR fading distribution. Deriving the PDF of ${\left\| {{{\mathbf{h}}_F}} \right\|^2}$ is equivalent to deriving the PDF of square of two-fold sum of product of FTR RVs, which is extremely difficult, if not impossible, to be obtained. Fortunately, the distribution of the sum of FTR random variables (RVs) can be accurately approximated by the single FTR distribution by following the method proposed in \cite{zhang2020Performance}, i.e., {\small $ \sum\limits_{m = 1}^M {\left| {g_1^{\left( {m,\ell } \right)}} \right|}  \approx \left| {g_1^{\left( \ell  \right)}} \right| $}. Thus, with the optimal phase shift design, we have
	{\small \begin{align}\label{jiangwei}
	{\left\| {{{\mathbf{h}}_F}} \right\|^2} &\!\approx\! {\left( {\sum\limits_{m = 1}^M {\sum\limits_{\ell  = 1}^L {\left| {g_2^{\left( \ell  \right)}} \right|\left| {g_1^{\left( {m,\ell } \right)}} \right|} } } \right)^2} \!= \!{\left(\! {\sum\limits_{\ell  = 1}^L {\left| {g_2^{\left( \ell  \right)}} \right|} \sum\limits_{m = 1}^M {\left| {g_1^{\left( {m,\ell } \right)}} \right|} } \!\right)^2}
	\notag\\
	& \approx \!{\left( {\sum\limits_{\ell  = 1}^L {\left| {g_2^{\left( \ell  \right)}} \right|} \left| {g_1^{\left( \ell  \right)}} \right|} \right)^2},
	\end{align}}\noindent
	where the two-fold summation is reduced to a one-fold summation, thus giving the possibility of further performance analysis. Note that \eqref{jiangwei} is also the exact small-scale fading expression when considering a single antenna transmitter. In the following, we analyze the SNR expressions in two cases.

	\subsubsection{Non-Ideal RF Chains}
	With the non-ideal RF chains, the instantaneous SNDR can be expressed as
	{\small \begin{equation}\label{gammaN}
	{\gamma _N} = \frac{{{{\left\| {{h_F}} \right\|}^2}{{\left| {{h_P}} \right|}^2}{{\left| {{h_L}} \right|}^2}P}}{{{\kappa ^2}{{\left\| {{h_F}} \right\|}^2}{{\left| {{h_P}} \right|}^2}{{\left| {{h_L}} \right|}^2}P + {N_o}}},
	\end{equation}}\noindent
	where $\kappa^{2} \triangleq \kappa_{S}^{2}+\kappa_{D}^{2}$. 
	\subsubsection{Ideal RF Chains}
	Let us consider the case of ideal RF chains. Using \eqref{gammaN} and letting ${\kappa =0}$, we can express the instantaneous SNR of the received signal as
	{\small \begin{equation}\label{gammaidcf}
	{\gamma _I} = \frac{{{ {{\left\| {{h_F}} \right\|}^2}{{\left| {{h_P}} \right|}^2}}{{\left| {{h_L}} \right|}^2}P}}{{{N_o}}}.
	\end{equation}}\noindent
	
	\section{Phase Shift Design Under Discrete Phase Shift Constraints }\label{afefaegva}
	In this section, we propose an SIO algorithm to obtain the optimal design of RIS's reflector array, ${\bf{\Phi}}$. With the optimal phase shift design of RIS, we can theoretically obtain the maximum of SNDR. In practical systems, ideal continuous phase-shifts are hard to implement because of hardware limitation. Thus, we consider the discrete phase-shifts design, which is more energy efficient \cite{wu2020beamforming}.
	
	\subsection{Optimal Phase Shift Design of RIS's Reflecting Elements}
	\subsubsection{Motivations for proposing the SIO algorithm}
	There are several challenges for the existing RIS phase shift design algorithms, which can be summarized as follows:
	\begin{itemize}
	\item When the number of reflecting elements at the RIS, $L$, is large, most of the existing algorithms \cite{du2020millimeter, wu2019intelligentsda,wu2020beamforming,feng2020deep} require a lot time to converge, which motivates us to propose an effective algorithm for practical RIS-aided systems. The running time should not be greatly effected by $L$.
	\item Eliminating the channel phases is difficult because the RIS is assumed to be passive. Although a novel and simple algorithm based on the binary search tree was proposed to obtain ${\bf{\Phi}}_{\rm opt}$ without the channel phases information \cite{du2020millimeter}, the algorithm proposed in \cite{du2020millimeter} is based on an assumption that the phase-shifts is continuous at each reflecting element, which is practically difficult to realize due to the hardware limitation of the RIS.
	\end{itemize}
	
	Combining the above discussion, we design an SIO algorithm to find ${\bf{\Phi}}_{\rm opt}$ for the RIS with finite-level phase-shifters without the channel phases information, inspired from Particle Swarm Optimization (PSO) and evolution algorithms. 
	
	\subsubsection{Introduction of the traditional PSO algorithm}
	For the traditional PSO algorithm, $n$ particles move in an $L$-dimensional search space to optimize a fitness function \cite{cheng2018a}. In a real RIS-aided THz communications system, the value of fitness function can be obtained by measuring the expectation of the amplitude of the received signal, $E_{\rm re}$, by averaging the received signal strength in the time domain. First, the particles are placed randomly in the search space. Then, for a predefined number of iterations, each particle evaluates fitness and moves to a new location based on the history of particles' own best previous location and other particles' best positions. Thus, the PSO algorithm consists of a swarm of $n$ particles, and the position of each particle represents a possible solution of the fitness function. The velocity $v_{i,j}$ and position $x_{i,j}$ of the $d_{\rm th}$ dimension of the $i_{\rm th}$ particle are updated as follows:
	{\small \begin{align}
	{v_{id}}\left( {k + 1} \right) = &\omega{v_{id}}\left( k \right) + {c_1}{R_1}\left( {{p_{id}}\left( k \right) - {x_{id}}\left( k \right)} \right)
	\notag\\&
	 + {c_2}{R_2}\left( {{p_{gd}}\left( k \right) - {x_{id}}\left( k \right)} \right),
	\end{align}}\noindent
	{\small \begin{equation}
	{x_{id}}\left( {k + 1} \right) = {x_{id}}\left( k \right) + {v_{id}}\left( {k + 1} \right),
	\end{equation}}\noindent
	where $i = 1,2, \ldots ,m$, $d = 1,2, \ldots ,L $, $\omega $ is the inertia weight, $c_1$ and $c_2$ are the acceleration constants, $R_1$ and $R_2$ are uniformly distributed random numbers in $[0, 1]$, $ {P_i} = \left( {{p_{i1}},{p_{i2}} \ldots ,{p_{iL}}} \right) $ is the best previous position of the $i_{\rm th}$ particle, and $ {P_g} = \left( {{p_{g1}},{p_{g2}} \ldots ,{p_{gL}}} \right) $ is the best position among all the particles.
	
	\subsubsection{Design of the SIO algorithm}
	The main disadvantage of the PSO algorithm is that it is difficult to balance local and global search and keep the diversity of the population. Thus, by combining the modified PSO algorithm \cite{clerc2002the,yoshida2000a} with evolution algorithms, we present the SIO Algorithm \ref{1}, which maintains the optimal balance between exploration and exploitation, and reduces computation time, to find ${\bf{\Phi}}_{\rm opt}$. In the following, we introduce the SIO's working mechanism and the important parameters: 
	\begin{itemize}
	\item Each particle is given a periodic value, ${\rm Pe}$, whose initial value is $0$. When the particle updates to a better state, the periodic value resets to $0$, and otherwise, the value increases by $1$. We set the size of swarm reaches as $n_{\rm begin}$. As the searching time increases, the particle that has the largest periodic value will be deleted, until the size of the swarm reaches $n_{\rm end}$. We can set $n_{\rm begin}=30$ and $n_{\rm end}=10$ to reduce the computational resources and keep the diversity of the population.
	\item The inertia weight $\omega$ can significantly affect the global and local search-ability. By linearly decreasing $\omega$ from a relatively larger value to a smaller value, the algorithm's global search-ability at the beginning and the local search-ability near the end of the search procedure can be efficiently enhanced \cite{juneja2016particle}. In experimental work (i.e., in Figs. \ref{ELement90}-\ref{TCOM}), $\omega$ is kept in between $0.9$ to $0.4$.
	\item Adding an exploitative component allows the algorithm to utilize local information to guide the search towards better regions in the search space. The accelerate constants, i.e., $c_3$, $c_1$ and $c_2$, represent the particle stochastic acceleration weights toward the best positions of the local, $i_{\rm th}$ particle and among all the particles. The system designer can set larger accelerate constants for fast search, or smaller accelerate constants to reduce oscillations of particles.
	\end{itemize}
	In contrast to the continuous phase-shifts, we consider the practical case where the RIS only has a finite number of discrete phase-shifts that are equally spaced in $[0, 2 \pi)$. The set of phase-shifts at each element is given by $\mathcal{O}=\{0, \Delta \theta, \ldots, \Delta \theta(K-1)\}$, where $K=\left\lceil {\frac{{{\rm{2}}\pi }}{{\Delta \theta }}} \right\rceil$. If we use $b$ bits to represent each of the levels, we can also express $K$ as $K=2^b$. Denote the best position of the local as $ {P_l} = \left( {{p_{l1}},{p_{l2}}, \ldots ,{p_{lD}}} \right) $, after finding the three best values, i.e., ${P_l}$, ${P_i}$, and ${P_g}$, the velocities and positions are updated with the following formulas:
	{\small \begin{equation}\label{sudu}
	{v_{id}}\left( {k + 1} \right) = \left[\left( \omega{v_{id}}\left( k \right) + \Delta v\left( k\right) \right) /\Delta\theta\right]\times \Delta\theta,
	\end{equation}}\noindent
	{\small \begin{align}\label{sudu2}
	\Delta v\left( k\right) = & {c_1}{R_1}\left( {{p_{id}}\left( k \right) - {x_{id}}\left( k \right)} \right) + {c_2}{R_2}\left( {{p_{gd}}\left( k \right) - {x_{id}}\left( k \right)} \right) 
	\notag\\
	&+ {c_3}{R_3}\left( {{p_{ld}}\left( k \right) - {x_{id}}\left( k \right)} \right),
	\end{align}}\noindent
	{\small \begin{equation}\label{weizhi}
	{x_{id}}\left( {k + 1} \right) = {x_{id}}\left( k \right)+ {v_{id}}\left( {k + 1} \right),
	\end{equation}}\noindent
	where $[\cdot]$ is the rounding function. From \eqref{sudu}-\eqref{weizhi}, we can observe that, in practical communications systems, we only need to measure the $E_{\rm re}$ in the time domain and adjust the RIS's phase shift with the help of Algorithm \ref{1}, and does not need any channel estimation to obtain ${\Phi _{{\rm opt}}}$.

	\begin{algorithm}[t]
	\caption{The swarm intelligence-based optimization algorithm to find ${\Phi _{{\rm opt}}}$ for the RIS with finite-level phase-shifters.}
	\label{1}
	\hspace*{0.02in} {\bf Input:} Input the number of elements on RIS: $L$, the minimum value of phase-shifts: $\Delta \theta$, maximum iterations: ${\rm Max}$, size of swarm: $n_{\rm begin}$ and $n_{\rm end}$, inertia weights: $\omega_{\rm begin}$ and $\omega_{\rm end}$, acceleration constants: $c_1$, $c_2$ and $c_3$, maximum of velocity: $v_{\rm max}$\\
	\hspace*{0.02in} {\bf Output:}
	The phase-shift variable ${\rm \varphi_{\ell}}$ $(\ell=1,\ldots,L)$ for each RIS element should be set.
	\begin{algorithmic}[1]
	\State Set iteration index $k=0$. Initialize the velocities and positions of $n$ particles, $\omega \leftarrow \omega_{\rm begin}-k/{\rm Max}*\left(\omega_{\rm begin}-\omega_{\rm end}\right)$.
	\For{Every praticle in swarm}
	\State Obtain corresponding expectation of the amplitude of the received signal, $E_{\rm re}$.
	\EndFor
	\State Update personal best position of each particle, global best position of the swarm, local best position and corresponding values.
	\While{$k<{\rm Max}$}
	\For{Every praticle}
	\State Obtain corresponding $E_{\rm re}$.
	\If{Current position is personal best position for this particle}
	\State Update its personal best position.
	\EndIf
	\If{Current value of $E_{\rm re}$ is smaller than the last value}
	\State Increase the periodic value: ${\rm Pe} \leftarrow {\rm Pe}+1$
	\EndIf
	\EndFor
	\State Update global best position, local best position and corresponding values
	\State Calculate the number of particles that will be deleted: $D \leftarrow n-\left[n_{\rm begin}-k/{\rm Max}*\left(n_{\rm begin}-n_{\rm end}\right)\right]$
	\If{$D>0$}
	\State Delete $D$ particles which have the largest periodic value, ${\rm Pe}$
	\EndIf
	\State Update the number of particles: $n \leftarrow n -D$
	\For{Every praticle}
	\State Update $v$ with the help of \eqref{sudu} and \eqref{sudu2}
	\If{$v>v_{\rm max}$}
	\State Update $v$: $v \leftarrow v_{\rm max}$
	\EndIf
	\State Update $x$ with the help of \eqref{weizhi}
	\EndFor
	\State Update the inertia weight: $\omega \leftarrow \omega_{\rm begin}-k/{\rm Max}*\left(\omega_{\rm begin}-\omega_{\rm end}\right)$
	\EndWhile
	\State \Return In $L$-dimensional search space, the global best position is ${\rm \varphi_{\ell}}$ $(\ell=1,\ldots,L)$		\end{algorithmic}
	\end{algorithm}
	
	\subsection{Convergence and Effectiveness Analysis}
	\subsubsection{Convergence Analysis of the SIO}
	To prove the convergence of our proposed SIO, we first derive the location iteration function. However, when $\Delta \theta \ne 0$, it is difficult to substitute \eqref{sudu} into \eqref{weizhi} because of the rounding function in \eqref{sudu}. To solve this problem, we rewrite \eqref{sudu} as
	{\small \begin{equation}\label{newsudu}
	{v_{id}}\left( {k + 1} \right) = \omega {v_{id}}\left( k \right) + \Delta v\left( k \right) + c\left( k \right),
	\end{equation}}\noindent
	where $ c\left( k \right) $ denotes the mismatch between ${\omega {v_{id}}\left( k \right) + \Delta v\left( k \right)} $ and RIS's practical discrete phases. We can obtain the value of $ c\left( k \right) $ by 
	{\small \begin{equation}
	c\left( k \right){\rm{ = }}\left[ {\left( {\omega {v_{id}}\left( k \right) + \Delta v\left( k \right)} \right)/\Delta \theta } \right] \times \Delta \theta  - \left( {\omega {v_{id}}\left( k \right) + \Delta v\left( k \right)} \right).
	\end{equation}}\noindent
	Thus, we have $  - \frac{1}{2}\Delta \theta  < c\left( k \right) < \frac{1}{2}\Delta \theta  $.	The velocity of each particle is updated as
	{\small \begin{equation}\label{suduweizhi}
	{v_{id}}\left( k+1 \right) = {x_{id}}\left( k+1 \right) - {x_{id}}\left( {k } \right) .
	\end{equation}}\noindent
	Let $T_i\triangleq c_iR_i$. Substituting \eqref{sudu2} and \eqref{newsudu} into \eqref{suduweizhi}, we can derive the location iteration function as
	{\small \begin{align}
	&{x_{id}}\left( {k + 1} \right) - \left( {1 - {T_1} - {T_2} - {T_3} + \omega } \right){x_{id}}\left( k \right) + \omega {x_{id}}\left( {k - 1} \right) \notag\\
	&= {T_1}{p_{id}}\left( k \right) + {T_2}{p_{gd}}\left( k \right) + {T_3}{p_{ld}}\left( k \right) + c\left( k \right).
	\end{align}}\noindent
	When the RIS's elements can adjust the phase-shifts continuously, we have $\Delta\theta = 0$ and $c\left( k\right) =0$. For the discrete case, the $c\left( k\right) =0$ can be regarded as a bounded noise. Note that the recurrence equation of SIO is linear non-homogeneous. To calculate the characteristic roots, we consider the secular equation as
	{\small 	\begin{equation}\label{secular}
	{x^2}-\left( {1-{T_1}-{T_2}-{T_3} + \omega } \right)x + \omega  = 0.
	\end{equation}}\noindent
	The discrete linear system that has such a secular equation has been well studied \cite{kan2012convergence}. Because the latent roots of \eqref{secular} can be derived easily, we can obtain the expression of $x(k)$ and prove that the SIO algorithm can converge to the optimal solution. Moreover, SIO suffers from a linear increase in the search complexity with an increase in the maximum iteration number, ${\rm Max}$. In addition, $E_{{\rm{re}}}$ is obtained $n$ times in each iteration. Thus, the complexity of SIO is given as $\mathcal{O}(n{\rm Max})$.
	
	\subsubsection{Effectiveness Analysis of the SIO}
	Let {\small $E_{{\rm{re}}} \triangleq {\mathbb E}\left[ {\left| {{h_L}} \right|\left| {{h_P}} \right|\left| {{h_F}} \right|} \right]$} and {\small $E_{{\rm{opt}}} \triangleq \left| {{h_L}} \right|{\mathbb E}\left[ {\left| {{h_P}} \right|} \right]{\mathbb E}\left[ {\left| {\sum\limits_{\ell  = 1}^L {{g_{1}^{\left(\ell\right)}}{g_{2}^{\left(\ell\right)}}} } \right|} \right]$}, where $E_{{\rm{re}}}$ is the fitness function in our SIO algorithm which can be easily measured in the time domain, and $E_{{\rm{opt}}}$ the maximum achievable $E_{{\rm{re}}}$, ${\mathbb E}\left[\cdot  \right]$ denotes the mathematical expectation.
	
	To investigate the effectiveness of the proposed phase optimization method, we first derive $E_{{\rm{opt}}}$ in the following theorem
	\begin{theorem}\label{aaa111t}
	The maximum achievable expectation of the amplitude of the received signal, $E_{{\rm{opt}}}$, can be derived as
	{\small \begin{equation}\label{eopt}
	E_{{\rm{opt}}} \! =\! \frac{{\left| {{h_L}} \right|{\gamma ^2}{A_o}}}{{{\gamma ^2} \!+\! 1}}\sum\limits_{\ell  = 1}^L \prod\limits_{\ell  = 1}^2 {\frac{{m_\ell ^{{m_\ell }}}}{{\Gamma\!\left(\! {{m_\ell }} \!\right)}}}\! \sum\limits_{{j_\ell } = 0}^{{N_T}} {\frac{{K_\ell ^{{j_\ell }}{d_{{\ell _{{j_\ell }}}}}{{\left( {2\sigma _\ell ^2} \right)}^{\frac{1}{2}}}}}{{{j_\ell }!\Gamma\!\left(\! {{j_\ell }\! + \!1} \!\right)}}} \Gamma\!\left(\! {\frac{3}{2} \!+ \!{j_\ell }} \!\right)\!\!.
	\end{equation}}\noindent
	\end{theorem}
	\begin{IEEEproof}
	Please refer to Appendix \ref{aaa111}.
	\end{IEEEproof}
	
	\begin{figure}[t]
	\centering
	\includegraphics[width=0.45\textwidth]{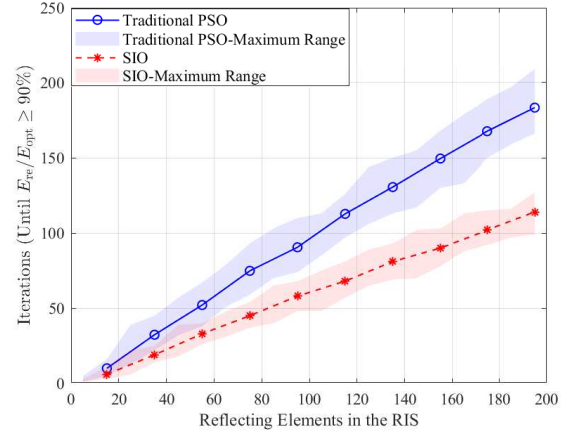}
	\caption{The number of iterations (until $E_{{\rm re}}/E_{{\rm opt}}\ge 90\%$) versus the number of RIS's elements, with $\Delta \theta=\pi/10$.}
	\label{ELement90}
\end{figure}
	\begin{figure}[t]
	\centering
	\includegraphics[width=0.45\textwidth]{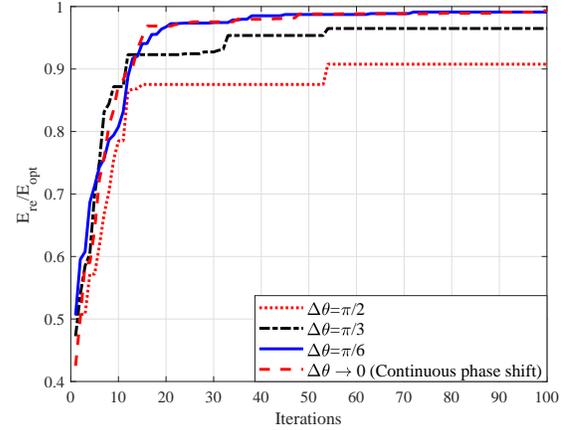}	
	\caption{The expectation error of the proposed SIO algorithm under different minimum values of phase-shifts versus the iterations, with $L=50$.}
	\label{FIGSIOPSO2}
\end{figure}
	\begin{figure}[t]
	\centering
	\includegraphics[width=0.45\textwidth]{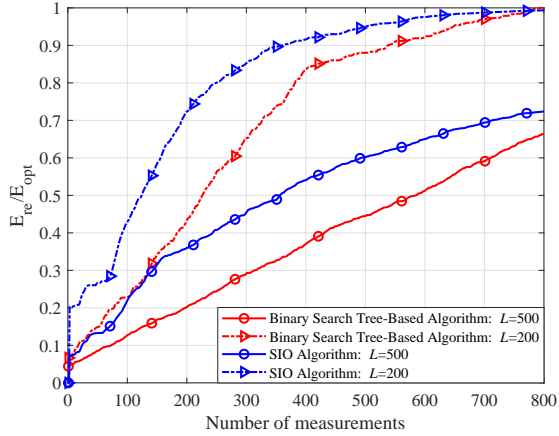}	
	\caption{The expectation error of the SIO algorithm and the algorithm proposed in \cite{du2020millimeter} versus the number of measurements of $E_{{\rm{re}}}$.}
	\label{TCOM}
	\end{figure}

	We use $E_{\rm re}/E_{\rm opt}$ to characterize the expectation accuracy and verify the convergence and effectiveness of the proposed algorithm. In Figs. \ref{ELement90} and \ref{FIGSIOPSO2}, we verify the convergence and effectiveness of the proposed algorithm for different iterations, by setting $P=40$ ${\rm dBW}$, $m_{\ell,1}=3$, $m_{\ell,2}=4$, $ {{2}}{\sigma _{\ell ,1}}\left( {1 + {K_{\ell ,1}}} \right) =20$ ${\rm dB}$, ${{2}}{\sigma _{\ell ,2}}\left( {1 + {K_{\ell ,2}}} \right) = 10$ ${\rm dB}$, $f=300$ ${\rm GHz}$, $d_1=30$ ${\rm m}$, $d_2=20$ ${\rm m}$, $N_0=1$ ${\rm dBW}$, $K_{\ell,1}=4$, $K_{\ell,2}=5$, $\Delta_{\ell,1}=0.7$, $\Delta_{\ell,2}=0.5$, $\kappa_S=\kappa_D=0.2$, $n_{\rm begin}=30$, $n_{\rm end}=10$, $c_1=1$, $c_2=2$, $c_3=2$, $\omega_{\rm begin}=0.9$, and $\omega_{\rm end}=0.4$. In Fig. \ref{ELement90}, the proposed SIO is compared with traditional PSO under the different number of reflecting elements, $L$. As it can be observed, although the numbers of iterations for both PSO and SIO increase linearly with the number of RIS elements, the increase rate when SIO is used is reduced by about $34\%$ compared to that when PSO is used, which means that the SIO is always superior to the traditional PSO at finding optimal phase-shifts. Moreover, the red and blue shaded ranges in Fig. \ref{ELement90} indicate the uncertainty in the actual operation of the algorithms. We can see that the fluctuation range of the number of iterations of the SIO (typically less than $25$) is smaller than that of the PSO. Figure \ref{FIGSIOPSO2} depicts $E_{\rm re}/E_{\rm opt}$ versus the amount of reflecting elements on the RIS, with different choices of $\Delta \theta$. Although the error of expectation increases as $\Delta \theta$ is increased, we can observe that using the RIS with large $\Delta \theta$, e.g., $\pi/4$, can achieve the accuracy close to that when using RIS with continuous phase-shifts. Furthermore, Fig. \ref{TCOM} shows that the proposed SIO algorithm is better than the binary search tree-based algorithm proposed in \cite{du2020millimeter}. Although both the proposed SIO algorithm and algorithm proposed in \cite{du2020millimeter} need to measure the $E_{{\rm{re}}}$, the number of reflecting elements only affects the number of measurements in the algorithm proposed in \cite{du2020millimeter}. Thus, especially when the RIS is equipped with a large number of elements, the SIO algorithm can achieve a higher $E_{\rm re}/E_{\rm opt}$ than that of the algorithm proposed in \cite{du2020millimeter} with the same number of measurements.
	
	\section{Exact Statistics of the End-to-End SNR and SNDR}\label{eagf34}
	Substituting $R = \sqrt \gamma $ into \eqref{PDFFTR}, we can easily obtain the PDF expression of FTR RVs. Let $ Y = \sum\limits_{\iota  = 1}^L {\prod\limits_{\ell  = 1}^N {{R_{\iota ,\ell }}} }  $ be the sum of product of $N$ FTR RVs, where $R_{\iota ,\ell } \sim {\cal F}{\cal T}{\cal R}\left( {{K_{\iota ,\ell } },{m_{\iota ,\ell } },{\Delta _{\iota ,\ell } },\sigma_{\iota ,\ell } ^2} \right)$ ($\iota=1,\ldots,L$ and $\ell=1,\ldots,N$), the exact PDF and CDF of $Y$ have been respectively derived in multivariate Fox's $H$-function \cite[eq. (A.1)]{mathai2009h} as \cite[eq. 11]{du2020millimeter} and \cite[eq. 12]{du2020millimeter}
	
	In Figs. \ref{ELement90}-\ref{TCOM}, we verify that Algorithm \ref{1} can be used to obtain the phase-shifts of RIS's elements which enable the received power tightly close to the optimal received power. Then, we present the instantaneous optimal SNR and SNDR for the case of ideal and non-ideal RF chains, respectively.
	
	\subsection{Non-Ideal RF Chains}
	To derive a generic analytical expression for the CDF of the instantaneous SNDR of non-ideal RF chains, we first present the following lemma with the help of \eqref{gammaN}.
	\begin{lemma}\label{jifenthem}
	Considering hardware impairments, we derive an integral representation for the PDF and CDF of ${\gamma _N}$ in \eqref{gammaN} assuming arbitrarily distributed $\left| {{h_F}} \right|$ and $\left| {{h_P}} \right|$ as
	{\small \begin{align}\label{jifencaSF}
	{f_{{\gamma _N}}}\left( x \right)& = \int_0^{{A_o}} {\frac{1}{y}} {f_{\left| {{h_F}} \right|}}\left( {\frac{{\rm{1}}}{y}\sqrt {\frac{{x{N_o}}}{{{{\left| {{h_L}} \right|}^2}P\left( {1 - x{\kappa ^2}} \right)}}} } \right){f_{\left| {{h_P}} \right|}}(y){\rm{d}}y
	\notag \\& \times
	\frac{1}{{2\sqrt {x{{\left( {1 - x{\kappa ^2}} \right)}^{\rm{3}}}\frac{{{{\left| {{h_L}} \right|}^2}P}}{{{N_o}}}} }},
	\end{align}}\noindent
	{\small \begin{equation}\label{jifencaSF2}
	{F_{{\gamma _N}}}\left( x \right) = \int_0^{{A_o}} {{F_{\left| {{h_F}} \right|}}} \left( {\frac{{\rm{1}}}{y}\sqrt {\frac{{x{N_o}}}{{{{\left| {{h_L}} \right|}^2}P\left( {1 - x{\kappa ^2}} \right)}}} } \right){f_{\left| {{h_P}} \right|}}(y){\rm{d}}y.
	\end{equation}}\noindent
	\begin{IEEEproof}
	Please refer to Appendix \ref{jifenappen}.
	\end{IEEEproof}
	\end{lemma}
	With the help of Lemma \ref{jifenthem}, for special distributed $\left| {{h_F}} \right|$ and $\left| {{h_P}} \right|$ in this paper, the statistics of instantaneous SNDR can be obtained as the following theorem.
	\begin{them}\label{hfpthem}
	The PDF and CDF of ${{\gamma _N}}$ can be derived as \eqref{GAMMANPDF} and \eqref{GAMMANCDF}, shown at the bottom of the next page, 	where $ \left\{ {\left( {{a_n}} \right)} \right\}_1^N = \left( {{a_1}} \right), \ldots ,\left( {{a_N}} \right) $, and ${\upsilon _{{h_{FP}},\ell }} = \left\{ {\left( { - {j_{\ell ,n}},0.5} \right)} \right\}_1^{\rm{2}},\left( {1 - {\gamma ^2},1} \right) $ $(\ell=1,\ldots,L)$.
	
		%----------------------------------------------------------------------------------------------------------------------------
	%----------------------------------------------------------------------------------------------------------------------------
	\newcounter{mycount523}
	\begin{figure*}[b]
		\normalsize
		\setcounter{mycount523}{\value{equation}}
		\hrulefill
		\vspace*{4pt}
	{\small \begin{align}\label{GAMMANPDF}
			&{f_{{\gamma _N}}}\left( x \right) =\sum\limits_{{j_{1,1}}, \ldots ,{j_{L,1}} = 0}^{N_T}     \sum\limits_{{j_{1,2}}, \ldots ,{j_{L,2}} = 0}^{N_T}  {\prod\limits_{\iota  = 1}^L {\prod\limits_{\ell  = 1}^2 {\frac{{{K_{\iota ,\ell }}^{{j_{\iota ,\ell }}}{d_{\iota ,\ell }}_{{j_{\iota ,\ell }}}}}{{{j_{\iota ,\ell }}!}}\frac{{{m_{\iota ,\ell }}^{{m_{\iota ,\ell }}}}}{{\Gamma ({m_{\iota ,\ell }})}}\frac{{{\gamma ^2}}}{{\Gamma ({j_{\iota ,\ell }} + 1)}}\frac{1}{{2x\left( {1 - x{\kappa ^2}} \right)}}} } }
			\notag\\&\times\!\!
			H_{1,0:3,{{2}}; \ldots ;3{{,2}}}^{0,0:1,3; \ldots ;1,3}\left(\!\!\! {\left. {\begin{array}{*{20}{c}}
						{{A_{{0}}}\sqrt {\frac{{{{\left| {{h_L}} \right|}^2}P\left( {1 - x{\kappa ^2}} \right)}}{{x{N_o}}}} \prod\limits_{\ell  = 1}^{{2}} {\left( {\sqrt 2 {\sigma _{1,\ell }}} \right)} }\\
						\vdots \\
						{{A_{{0}}}\sqrt {\frac{{{{\left| {{h_L}} \right|}^2}P\left( {1 - x{\kappa ^2}} \right)}}{{x{N_o}}}} \prod\limits_{\ell  = 1}^{{2}} {\left( {\sqrt 2 {\sigma _{L,\ell }}} \right)} }
				\end{array}}\! \right|\begin{array}{*{20}{c}}
					{\left( {0;1, \ldots ,1} \right):{\upsilon _{{h_{FP}},1}}; \ldots ;{\upsilon _{{h_{FP}},L}}}\\
					{ - :\left( {0,1} \right),\left( { - {\gamma ^2},1} \right); \ldots ;\left( {0,1} \right),\left( { - {\gamma ^2},1} \right)}
			\end{array}}\!\!\!\! \right)
	\end{align}}
	{\small \begin{align}\label{GAMMANCDF}
			&{F_{{\gamma _N}}}\left( x \right) = \sum\limits_{{j_{1,1}}, \ldots ,{j_{L,1}} = 0}^{N_T}     \sum\limits_{{j_{1,2}}, \ldots ,{j_{L,2}} = 0}^{N_T}  {\prod\limits_{\iota  = 1}^L {\prod\limits_{\ell  = 1}^2 {\frac{{{K_{\iota ,\ell }}^{{j_{\iota ,\ell }}}{d_{\iota ,\ell }}_{{j_{\iota ,\ell }}}}}{{{j_{\iota ,\ell }}!}}\frac{{{m_{\iota ,\ell }}^{{m_{\iota ,\ell }}}}}{{\Gamma ({m_{\iota ,\ell }})}}\frac{1}{{\Gamma ({j_{\iota ,\ell }} + 1)}}} } } {\gamma ^2}
			\notag\\&\times\!\!
			H_{1,0:3,{{2}}; \ldots ;3{{,2}}}^{0,0:1,3; \ldots ;1,3}\left(\!\!\! {\left. {\begin{array}{*{20}{c}}
						{{A_{{0}}}\sqrt {\frac{{{{\left| {{h_L}} \right|}^2}P\left( {1 - x{\kappa ^2}} \right)}}{{x{N_o}}}} \prod\limits_{\ell  = 1}^{{2}} {\left( {\sqrt 2 {\sigma _{1,\ell }}} \right)} }\\
						\vdots \\
						{{A_{{0}}}\sqrt {\frac{{{{\left| {{h_L}} \right|}^2}P\left( {1 - x{\kappa ^2}} \right)}}{{x{N_o}}}} \prod\limits_{\ell  = 1}^{{2}} {\left( {\sqrt 2 {\sigma _{L,\ell }}} \right)} }
				\end{array}} \right|\begin{array}{*{20}{c}}
					{\left( {1;1, \ldots ,1} \right):{\upsilon _{{h_{FP}},1}}; \ldots ;{\upsilon _{{h_{FP}},L}}}\\
					{ - :\left( {0,1} \right),\left( { - {\gamma ^2},1} \right); \ldots ;\left( {0,1} \right),\left( { - {\gamma ^2},1} \right)}
			\end{array}}\!\!\right)
	\end{align}}
		\setcounter{equation}{\value{mycount523}}
	\end{figure*}
	\addtocounter{equation}{2}
	%----------------------------------------------------------------------------------------------------------------------------
	%----------------------------------------------------------------------------------------------------------------------------
	\begin{IEEEproof}
	Please refer to Appendix \ref{hfpappen}.
	\end{IEEEproof}
	\end{them}
	
	\begin{rem}
	The multivariate Fox's $H$-function cannot be evaluated directly by popular mathematical packages such as MATLAB or Mathematica. However, many implementations have been proposed \cite{vega2019statistics,alhennawi2015closed}. In the following, we use an efficient and accurate Python implementation \cite{alhennawi2015closed} which can provide accurate results of multivariate Fox's $H$-function quickly by estimating the multivariate integrals with simple rectangle quadrature. For example, the execution time for calculating ternary Fox's $H$-function is less than a few seconds.
	\end{rem}
	
	\subsection{Ideal RF Chains}
	\begin{them}
	The PDF and CDF of ${{\gamma _I}}$ can be derived as \eqref{qy2uh4i312} and \eqref{GAMMAICDF}, shown at the bottom of the next page.
	%----------------------------------------------------------------------------------------------------------------------------
	%----------------------------------------------------------------------------------------------------------------------------
	\newcounter{mycount52}
	\begin{figure*}[b]
		\normalsize
		\setcounter{mycount52}{\value{equation}}
		\hrulefill
		\vspace*{4pt}
	{\small \begin{align}\label{qy2uh4i312}
			&{f_{{\gamma _I}}}\left( x \right) =\sum\limits_{{j_{1,1}}, \ldots ,{j_{L,1}} = 0}^{N_T}    \sum\limits_{{j_{1,2}}, \ldots ,{j_{L,2}} = 0}^{N_T}  {\prod\limits_{\iota  = 1}^L {\prod\limits_{\ell  = 1}^2 {\frac{{{K_{\iota ,\ell }}^{{j_{\iota ,\ell }}}{d_{\iota ,\ell }}_{{j_{\iota ,\ell }}}}}{{{j_{\iota ,\ell }}!}}\frac{{{m_{\iota ,\ell }}^{{m_{\iota ,\ell }}}}}{{\Gamma ({m_{\iota ,\ell }})}}\frac{{{\gamma ^2}}}{{\Gamma ({j_{\iota ,\ell }} + 1)}}\frac{{1}}{2 x}} } }
			\notag\\&\times
			H_{1,0:3,{{2}}; \ldots ;3{\rm{,2}}}^{0,0:1,3; \ldots ;1,3}\left( {\left. {\begin{array}{*{20}{c}}
						{{A_{{0}}}\sqrt {\frac{{{{\left| {{h_L}} \right|}^2}P}}{{x{N_o}}}} \prod\limits_{\ell  = 1}^{{2}} {\left( {\sqrt 2 {\sigma _{1,\ell }}} \right)} }\\
						\vdots \\
						{{A_{{0}}}\sqrt {\frac{{{{\left| {{h_L}} \right|}^2}P}}{{x{N_o}}}} \prod\limits_{\ell  = 1}^{{2}} {\left( {\sqrt 2 {\sigma _{L,\ell }}} \right)} }
				\end{array}} \right|\begin{array}{*{20}{c}}
					{\left( {0;1, \ldots ,1} \right):{\upsilon _{{h_{FP}},1}}; \ldots ;{\upsilon _{{h_{FP}},L}}}\\
					{ - :\left( {0,1} \right),\left( { - {\gamma ^2},1} \right); \ldots ;\left( {0,1} \right),\left( { - {\gamma ^2},1} \right)}
			\end{array}} \right)
	\end{align}}\noindent
	{\small \begin{align}\label{GAMMAICDF}
			&{F_{{\gamma _I}}}\left( x \right) = \sum\limits_{{j_{1,1}}, \ldots ,{j_{L,1}} = 0}^{N_T}    \sum\limits_{{j_{1,2}}, \ldots ,{j_{L,2}} = 0}^{N_T}  {\prod\limits_{\iota  = 1}^L {\prod\limits_{\ell  = 1}^2 {\frac{{{K_{\iota ,\ell }}^{{j_{\iota ,\ell }}}{d_{\iota ,\ell }}_{{j_{\iota ,\ell }}}}}{{{j_{\iota ,\ell }}!}}\frac{{{m_{\iota ,\ell }}^{{m_{\iota ,\ell }}}}}{{\Gamma ({m_{\iota ,\ell }})}}\frac{1}{{\Gamma ({j_{\iota ,\ell }} + 1)}}} } } {\gamma ^2}
			\notag\\&\times
			H_{1,0:3,{{2}}; \ldots ;3{{,2}}}^{0,0:1,3; \ldots ;1,3}\left( {\left. {\begin{array}{*{20}{c}}
						{{A_{\rm{0}}}\sqrt {\frac{{{{\left| {{h_L}} \right|}^2}P}}{{x{N_o}}}} \prod\limits_{\ell  = 1}^{{2}} {\left( {\sqrt 2 {\sigma _{1,\ell }}} \right)} }\\
						\vdots \\
						{{A_{{0}}}\sqrt {\frac{{{{\left| {{h_L}} \right|}^2}P}}{{x{N_o}}}} \prod\limits_{\ell  = 1}^{{2}} {\left( {\sqrt 2 {\sigma _{L,\ell }}} \right)} }
				\end{array}} \right|\begin{array}{*{20}{c}}
					{\left( {1;1, \ldots ,1} \right):{\upsilon _{{h_{FP}},1}}; \ldots ;{\upsilon _{{h_{FP}},L}}}\\
					{ - :\left( {0,1} \right),\left( { - {\gamma ^2},1} \right); \ldots ;\left( {0,1} \right),\left( { - {\gamma ^2},1} \right)}
			\end{array}} \right)
	\end{align}}\noindent
		\setcounter{equation}{\value{mycount52}}
	\end{figure*}
	\addtocounter{equation}{2}
	%----------------------------------------------------------------------------------------------------------------------------
	%----------------------------------------------------------------------------------------------------------------------------
	
	\begin{IEEEproof}
	Substituting $\kappa^{2}=0$ into \eqref{GAMMANPDF} and \eqref{GAMMANCDF}, we obtain the PDF and CDF of ${{\gamma _N}}$ after performing some mathematical manipulations, which completes the proof.
	\end{IEEEproof}
	\end{them}
	The statistics of $\gamma_I$ and $\gamma_I$ that we derived in Theorem 2 and Theorem 3 can be used to obtain exact closed-form expressions for key performance metrics.
	\section{Asymptotic Analysis}\label{casfae341}
	Although the previously derived analytical results are obtained in closed-form, they may not provide many insights as to the factors affecting system performance. In the following, the asymptotic performance analysis that becomes tight in high-SNDR and $L$ regimes is presented.
	\subsection{Asymptotic analysis in the high SNDR regime}
	\begin{prop}\label{Prop1}
	A simple approximation of the CDF of $\gamma _N$ in the high-SNDR regime can be derived as
	{\small \begin{align}\label{afjejo}
	&F_{{\gamma _N}}^{HS}\left( x \right) \!=\!\!\!\sum\limits_{{j_{1,1}}, \ldots ,{j_{L,1}} = 0}^{{N_T}} {\sum\limits_{{j_{1,2}}, \ldots ,{j_{L,2}} = 0}^{{N_T}} {\prod\limits_{\iota  = 1}^L {\prod\limits_{\ell  = 1}^2 {\frac{{{K_{\iota ,\ell }}^{{j_{\iota ,\ell }}}{d_{\iota ,\ell }}_{{j_{\iota ,\ell }}}{m_{\iota ,\ell }}^{{m_{\iota ,\ell }}}{\gamma ^2}}}{{{j_{\iota ,\ell }}!\Gamma\!\left(\! {{m_{\iota ,\ell }}} \!\right)\!\Gamma\!\left(\! {{j_{\iota ,\ell }} \!+\! 1} \!\right)}}} } } } 
	\notag\\&\times\!
	\prod\limits_{\iota  = 1}^L \frac{{\mathop {\lim }\limits_{{\varsigma _\iota } \to  - {\varpi _\iota }} \left( {{\varsigma _\iota } + {\varpi _\iota }} \right)\prod\limits_{\ell  = 1}^2 {\Gamma\! \left( {1 \!+\! {j_{\iota ,\ell }}\! -\! \frac{1}{2}{\varpi _\iota }} \right)} \Gamma\! \left( {{\varpi _\iota }} \right)\Gamma\! \left( {{\gamma ^2} \!-\!{\varpi _\iota }} \right)}}{{\Gamma\! \left( {1\! +\! \sum\limits_{\iota {\text{ = 1}}}^L {{\varpi _\iota }} } \right)\! \Gamma \!\left( {{\gamma ^2} \!-\! {\varpi _\iota } \!+ \!1} \right)}}
	\notag\\&\times\!
	{{\left(\!\! {{A_{\text{0}}}\sqrt {\frac{{{{\left| {{h_L}} \right|}^2}P\left( {1 \!- \!x{\kappa ^2}} \right)}}{{x{N_o}}}} \prod\limits_{\ell  = 1}^2 {\left( {\sqrt 2 {\sigma _{\iota ,\ell }}} \right)} } \!\right)}^{ - {\varpi _\iota }}},
	\end{align}}\noindent
	where $ {\varpi _\iota } = \min \left\{ {{\gamma ^2},2 + 2{j_{\iota ,1}},2 + 2{j_{\iota ,2}}} \right\} $, and
	{\small \begin{align}
	&\mathop {\lim }\limits_{{\varsigma _\iota } \to  - {\varpi _\iota }} \left( {{\varsigma _\iota } + {\varpi _\iota }} \right)\prod\limits_{\ell  = 1}^2 {\Gamma \left( {1 + {j_{\iota ,\ell }} - \frac{1}{2}{\varpi _\iota }} \right)} \Gamma \left( {{\varpi _\iota }} \right)\Gamma\!\left( {{\gamma ^2} - {\varpi _\iota }} \right)
	\notag\\& =
	\left\{ \begin{gathered}
	\prod\limits_{\ell  = 1}^2 {\Gamma\!\left( {1 + {j_{\iota ,\ell }} - \frac{1}{2}{\gamma ^2}} \right)} \Gamma\!\left( {{\gamma ^2}} \right),{\varpi _\iota } = {\gamma ^2} \hfill \\
	\Gamma\!\left( {{j_{\iota ,2}} - {j_{\iota ,1}}} \right)\Gamma\!\left( {2 + 2{j_{\iota ,1}}} \right)\Gamma\!\left( {{\gamma ^2} - 2 - 2{j_{\iota ,1}}} \right),{\varpi _\iota } = 2 + 2{j_{\iota ,1}} \hfill \\
	\Gamma\!\left( {{j_{\iota ,1}} - {j_{\iota ,2}}} \right)\Gamma\!\left( {2 + 2{j_{\iota ,2}}} \right)\Gamma\!\left( {{\gamma ^2} - 2 - 2{j_{\iota ,2}}} \right),{\varpi _\iota } = 2 + 2{j_{\iota ,2}} \hfill \\
	\end{gathered}  \right.
	\end{align}}\noindent
	\end{prop}
	\begin{IEEEproof}
	With the help of the definition of multivariate Fox's $H$-function \cite[eq. (A-1)]{mathai2009h}, eq. \eqref{GAMMANCDF} can be expressed in terms of multiple Mellin-Barnes type contour integrals. When $SNDR\to \infty$, we have $P\to \infty$. The Mellin-Barnes integrals can be approximated by evaluating the residue at the minimum pole on the left-hand side of $d{\varsigma _\iota }$ as \cite[Theorem 1.11]{mathai2009h}. Therefore, the expression in \eqref{afjejo} is derived, which completes the proof.
	\end{IEEEproof}
	Note that the CDF of ${{\gamma _I}}$ in the high-SNDR regime can be derived by substituting $\kappa^{2}=0$ into \eqref{afjejo}.
	\subsection{Asymptotic analysis in the high $L$ regime}
	Considering the RIS typically has hundreds of elements, we next derive closed-form statistics of the RV $ {{h_F}} = \sum\limits_{\iota  = 1}^L {\prod\limits_{\ell  = 1}^2 {{R_{\iota ,\ell}}}}$ in high-SNDR and $L$ regimes. Employing central limit theorem (CLT) as \cite{selimis2021on}, we have $ {{h_F}} \sim \mathcal{N}\left( {L{\mathbb E}\left[ {h_F} \right],L{\mathbb V}\left[ {h_F} \right]} \right) $, where ${\mathbb V}(\cdot )$ denotes the variance of an RV. Considering that ${\mathbb V}(\cdot )$=${\mathbb E}(\cdot^2)$-${\mathbb E}^2(\cdot )$, we can obtain the values of ${\mathbb E}\left[ {h_F} \right]$ and ${\mathbb V}\left[ {{h_F}} \right]$ with the help of \cite[eq. (6)]{du2020millimeter}. Therefore, the CDF of ${h_F}$ in the high $L$ regime can be expressed as \cite{laplace1995theorie}
	{\small \begin{equation}\label{sfaeafa}
	F_{{h_F}}^{HL}\!\left( x \right) \!= \!\frac{1}{2}\!\left(\! {1 \!+\! {\rm erf}\left(\! {\frac{{x - L{\mathbb E}\left[ {h_F} \right]}}{{\sqrt {2L{\mathbb V}\left[ {h_F} \right]} }}} \!\right)} \!\right) \! =\!\frac{1}{2} { {\rm erfc}\!\left(\! -{\frac{{x - L{\mathbb E}\left[ {h_F} \right]}}{{\sqrt {2L{\mathbb V}\left[ {h_F} \right]} }}}\!\right)},
	\end{equation}}\noindent
	where ${\rm erf}(\cdot)$ is the error function \cite[eq. (8.250.1)]{gradshteyn2007}, and ${\rm erfc}(\cdot)$ is the complementary error function \cite[eq. (8.250.4)]{gradshteyn2007}. When $L \to \infty$, the CDF of $\gamma _N$ can be derived with the help of \eqref{sfaeafa}, by following the similar method as shown in Appendix \ref{hfpappen}. However, note that ${\rm erf}\left( \cdot\right) $ or ${\rm erfc}\left( \cdot\right) $ in \eqref{sfaeafa} is a non-elementary function, which is defined in the form of a definite integral. Thus, it is hard to obtain new insights. Therefore, we use a five-first-moment (5FM) based method to obtain the approximate CDF expression. 
	\begin{prop}\label{newprop}
	An approximation of the CDF of $\gamma _N$ in the high-$L$ regime can be expressed as
	{\small \begin{equation}\label{qwerlk}
	F_{{\gamma _N}}^{HL}\left( x \right) = {a_1}{a_2}{\gamma ^2}G_{3,4}^{3,1}\left( {\left. {\frac{\Upsilon }{{{A_o}{a_2}}}} \right|\begin{array}{*{20}{c}}
	{1,{a_3} + 1,{\gamma ^2} + 1}\\
	{{a_4} + 1,{a_5} + 1,{\gamma ^2},0}
	\end{array}} \right),
	\end{equation}}\noindent
	where {\small $\Upsilon \triangleq \sqrt {\frac{{x{N_o}}}{{{{\left| {{h_L}} \right|}^2}P\left( {1 - x{\kappa ^2}} \right)}}}$}, {\small $G_{ \cdot , \cdot }^{ \cdot , \cdot }\left( { \cdot \left| \cdot \right.} \right)$} is the Meijer's $G$ function \cite[eq. (9.30)]{gradshteyn2007}, $ {\Omega _i} $ denotes the $i_{\rm th}$ moment of $h_F$, {\small $ {\varphi _i} = {\Omega _i}/{\Omega _{i - 1}} $}, {\small $ {a_3} = \left( {4{\varphi _4} - 9{\varphi _3} + 6{\varphi _2} - {\Omega _1}} \right){\rm{/}}\left( {3{\varphi _3} - {\varphi _4} - 3{\varphi _2} + {\Omega _1}} \right) $}, {\small $ {a_2} = \frac{{{a_3}}}{2}\left( {{\varphi _4} - 2{\varphi _3} + {\varphi _2}} \right) + 2{\varphi _4} - 3{\varphi _3} + {\varphi _2} $}, {\small $ {a_6} = \left( {{a_3}\left( {{\varphi _2} - {\Omega _1}} \right) + 2{\varphi _2} - {\Omega _1}} \right)/{a_2} - 3 $}, {\small $ {a_7} = \sqrt {{{\left( {{a_6} + 2} \right)}^2} - 4{\Omega _1}\left( {{a_3} + 1} \right)/{a_2}}  $}, {\small $ {a_4} = \left( {{a_6} + {a_7}} \right){\rm{/2}} $}, {\small $ {a_5} = \left( {{a_6} - {a_7}} \right){\rm{/2}} $}, and {\small $ {a_1} ={\Gamma \left( {{a_3} + 1} \right)/\left( {{a_2}\Gamma \left( {{a_4} + 1} \right)\Gamma \left( {{a_5} + 1} \right)} \right)} $}.
	\end{prop}
	\begin{IEEEproof}
	Please refer to Appendix \ref{newproof}.
	\end{IEEEproof}
	By substituting $\kappa^{2}=0$ in \eqref{qwerlk}, after performing some mathematical manipulations, we can obtain the CDF of ${{\gamma _I}}$ in the high $L$ regime.
	
	\subsection{Verification and Insights}
	\begin{figure}[t]
	\centering
	\includegraphics[width=0.45\textwidth]{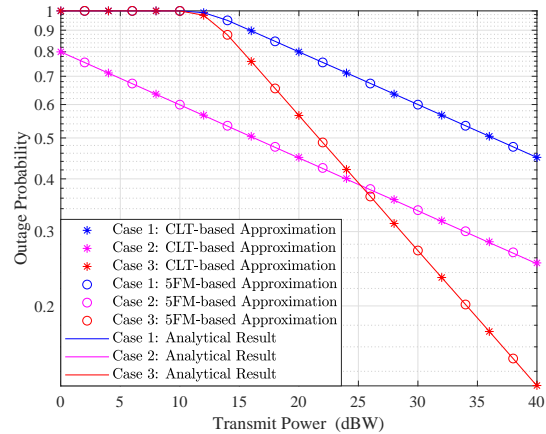}	
	\caption{Outage probability versus the transmit power with $L=20$, $A_o=0.01$, $K_{\ell,1}=5$, $K_{\ell,2}=6$, $m_{\ell,1}=5$, $m_{\ell,2}=7$, $ {{2}}{\sigma _{\ell ,1}}\left( {1 + {K_{\ell ,1}}} \right) = {{2}}{\sigma _{\ell ,2}}\left( {1 + {K_{\ell ,2}}} \right) = 10$ ${\rm dB}$, $\Delta_{\ell,1}=0.6$, $\Delta_{\ell,2}=0.4$, $N_0=1$ ${\rm dBW}$, $\gamma_{\rm th}=0.5$ ${\rm dB}$, and $\kappa_S=\kappa_D=0.1$ in three different cases, where case 1 is $\left| {{h_L}} \right|=0.1$ and $\gamma=0.5$, case 2 is $\left| {{h_L}} \right|=1$ and $\gamma=0.5$, and case 3 is $\left| {{h_L}} \right|=0.1$ and $\gamma=0.8$.}
	\label{VARIF}
	\end{figure}
	
	With the help of the approximate CDF expressions in Propositions \ref{Prop1} and \ref{newprop}, the difficulty of deriving performance metrics for RIS-aided THz communications systems can be avoided. For example, the outage probability $P$ can be defined as the probability that the received SNR per signal falls below a given threshold $\gamma_{\rm th}$, which can be directly obtained using the derived CDF expressions. Figure \ref{VARIF} shows the outage probability versus the transmit power. We can observe that the analytical results (\eqref{GAMMANCDF} in Theorem \ref{hfpthem}) match well with the high-$L$ approximate results obtained by the two methods, i.e., CLT and 5FM, in Proposition \ref{newprop} with $L=20$, which means that the high-$L$ approximate expressions are useful because the RIS can be equipped with hundreds of reflecting elements in the practical system. Note that when the number of reflecting elements is small, the CLT method is not accurate, but the 5FM method can be used because the parameter $ {\Omega _i} $, which denotes the $i_{\rm th}$ moment of small-scale fading coefficient, contains the impact of the number of elements. Furthermore, the accuracy of the approximate CDF expressions derived in Proposition \ref{Prop1} is verified in Fig. \ref{FIGOP} in the following numerical analysis section. We can observe that at high-transmitting power regime, \eqref{afjejo} approximates well the exact analytical expression.
	
	In addition to simplifying the performance analysis, we can observe several insights from the approximate expressions. For example, similar to the diversity order \cite{zheng2003diversity}, the high-SNR slope is a key performance indicator that can explicitly capture the impact of channel parameters on OP performance, which is defined as \cite{lozano2005high}
	{\small \begin{equation}\label{detf}
	{G_d} =  - \mathop {\lim }\limits_{P \to \infty } \frac{{\log \left( {O{P_{{\rm{appr}}}}} \right)}}{{\log \left( P \right)}}.
	\end{equation}}\noindent
	With the help of the accurate 5FM-based approximation \eqref{qwerlk}, we derive the simple approximate OP in high-SNDR regime as
	{\small \begin{align}\label{faegf12}
	&{{\rm OP}_{{\rm appr}}}\left( x \right) \! = \!\frac{{\mathop {\lim }\limits_{s \to {a_{\min }}}\! \left( {{a_{\min }} - s} \right){\Gamma ^{ - 1}}\left( {{\gamma ^2}\!+ \!1 \!-\! {a_{\min }}} \right)}}{{\Gamma\!\left( {{a_3} \!+ \!1 \!-\! {a_{\min }}} \right)\Gamma\!\left( {{a_{\min }} \!+ \!1} \right)}}{\left(\! {\frac{\Upsilon }{{{A_o}{a_2}}}} \!\right)^{{a_{\min }}}}
	\notag\\
	&\times {a_1}{a_2}{\gamma ^2}\Gamma \left( {{a_4} + 1 - s} \right)\Gamma \left( {{a_5} + 1 - s} \right)\Gamma \left( {{\gamma ^2} - s} \right)\Gamma \left( s \right),
	\end{align}}\noindent
	where {\small $ {a_{\min }} = \min \left\{ {{a_5} + 1,{a_4} + 1,{\gamma ^2}} \right\} $} and
	{\small \begin{align} 
&\mathop {\lim }\limits_{s \to {a_{\min }}} \left( {{a_{\min }} - s} \right)\Gamma \left( {{a_4} + 1 - s} \right)\Gamma \left( {{a_5} + 1 - s} \right)\Gamma \left( {{\gamma ^2} - s} \right)\Gamma \left( s \right)
	\notag\\&
	=\! \left\{\!\!\! {\begin{array}{*{20}{l}}
	{\Gamma\!\left( {{a_5} - {a_4}} \right)\Gamma\!\left( {{\gamma ^2} - {a_4} - 1} \right)\Gamma\!\left( {{a_4} \!+\! 1} \right),}&\!\!{{a_{\min }} = {a_4} + 1}\\
	{\Gamma\!\left( {{a_4} - {a_5}} \right)\Gamma\!\left( {{\gamma ^2} - {a_5} - 1} \right)\Gamma\!\left( {{a_5} \!+\! 1} \right),}&\!\!{{a_{\min }} = {a_5} + 1}\\
	{\Gamma\!\left( {{a_4} \!+\! 1 - {\gamma ^2}} \right)\Gamma\!\left( {{a_5} \!+ \!1 - {\gamma ^2}} \right)\Gamma\!\left( {{\gamma ^2}} \right),}&\!\!{{a_{\min }} = {\gamma ^2}}
	\end{array}} \right.
	\end{align}}\noindent
	\begin{IEEEproof}
	In the high-SNDR regime, e.g., when the transmit power is high, we have $\Upsilon \to 0$. The Mellin-Barnes integral over $\mathcal{L}_s$ can be approximated by evaluating the residue at the minimum pole on the left-hand side of $s$ as \cite[Theorem 1.11]{mathai2009h}. With the help of \cite[eq. (8.331.1)]{gradshteyn2007} and \cite[eq. (9.113)]{gradshteyn2007}, we can obtain \eqref{faegf12}, which completes the proof.	
	\end{IEEEproof}
	
	From \eqref{faegf12} and \eqref{detf}, we can obtain that $ {G_d} = {a_{\min}}/2 =\min \left\{ {{a_5} + 1,{a_4} + 1,{\gamma ^2}} \right\}/2$. Recall that $ {{a_4}} $ and $ {{a_5}} $ are determined by the moments of the small-scale fading coefficient, and ${\gamma ^2}$ is defined as the ratio of THz signal's equivalent beam-width and the variance of the pointing error displacement at Destination. Considering that ${G_d}$ indicates how much the OP decreases with the increase of the transmit power, we can conclude that when the pointing error is large, the key to improve ${G_d}$ lies in enhancing the beamforming to increase $\gamma$. when the pointing error is small, ${G_d}$ is mainly determined by the small-scale fading. This observation is verified in Fig. \ref{VARIF}: the blue and pink lines, i.e., case 1 and case 2, respectively, use the same value of ${G_d}$, so they have the same slope in the high SNDR region. However, in the simulation of the red line, i.e., case 3, the ${G_d}$ is increased from $0.5$ to $0.8$, so that the maximum slope is achieved.

	\section{Performance Analysis}\label{casfae34}
	\subsection{Outage Probability}
	The outage probability of considered THz communications can be obtained as ${P_{{\rm{out}}}} = P\left( {\gamma_\ell < {\gamma _{{\rm{th}}}}} \right){{ = }}F_{\gamma_\ell}\left( {{\gamma _{{\rm{th}}}}} \right)$, where $\ell  \in N,I$. For the case of non-ideal RF chains, the outage probability of the RIS-aided system can be directly obtained by using \eqref{GAMMANCDF}. Furthermore, we can obtain the outage probability for the case of ideal RF chains with the help of \eqref{GAMMAICDF}. Moreover, in high SNDR and $L$ regimes, the outage probability can be derived with the help of \eqref{afjejo} and \eqref{qwerlk}, respectively. Thus, we can observe that an RIS equipped with more reflecting elements will also achieve a low outage probability.
	\subsection{Ergodic Capacity}
	The ergodic capacity under optimal rate adaptation with constant transmit power is given by \cite[eq. (29)]{alouini1999capacity} as $C = {\mathbb E}\left[ {{{\log }_2}\left( {1 + {\gamma _\ell}} \right)} \right]$, where $\ell  \in N,I$.
	\subsubsection{Non-Ideal RF Chains}
	For the case of non-ideal RF chains, an exact expression of the ergodic capacity is hard to be derived due to the complexity of \eqref{GAMMANPDF}. However, the following proposition provides a closed-form upper bound of the ergodic capacity, which is more generic, and practically more useful and reliable. Moreover, in Fig. \ref{FIGEC}, we show that the derived capacity upper bound is close to the exact simulation results.
	
	\begin{prop}\label{ECpropapp}
	The effective capacity of the RIS-aid THz communications for the case of non-ideal RF chains is bounded by
	{\small \begin{equation}\label{afegnyt6r}
	C_{N} \le {\log _2}\left( {1 + \frac{{P{{\left| {{h_L}} \right|}^2}{\mathbb E}\left[ {{{\left| {{h_{FP}}} \right|}^2}} \right]}}{{P{\kappa ^2}{{\left| {{h_L}} \right|}^2}{\mathbb E}\left[ {{{\left| {{h_{FP}}} \right|}^2}} \right] + {N_o}}}} \right),
	\end{equation}}\noindent
	where ${\mathbb E}\left[ {{{\left| {{h_{FP}}} \right|}^2}} \right]$ is derived as \eqref{fagekjg}, shown at the bottom of the next page.

	%----------------------------------------------------------------------------------------------------------------------------
	%----------------------------------------------------------------------------------------------------------------------------
	\newcounter{mycount5}
	\begin{figure*}[b]
		\normalsize
		\setcounter{mycount5}{\value{equation}}
		\hrulefill
		\vspace*{4pt}
	{\small \begin{align}\label{fagekjg}
			&{\mathbb E}\left[ {{{\left| {{h_{FP}}} \right|}^2}} \right] = \sum\limits_{{j_{1,1}}, \ldots ,{j_{L,1}} = 0}^{N_T}  {\sum\limits_{{j_{1,{\rm{2}}}}, \ldots ,{j_{L,{\rm{2}}}} = 0}^{N_T}  {\prod\limits_{\iota  = 1}^L {\prod\limits_{\ell  = 1}^2 {\frac{{{K_{\iota ,\ell }}^{{j_{\iota ,\ell }}}{d_{\iota ,\ell }}_{{j_{\iota ,\ell }}}}}{{{j_{\iota ,\ell }}!}}\frac{{{m_{\iota ,\ell }}^{{m_{\iota ,\ell }}}}}{{\Gamma ({m_{\iota ,\ell }})}}\frac{{{\gamma ^2}}}{{{e^2}\Gamma ({j_{\iota ,\ell }} + 1)}}} } } }
			\notag\\&\times\!\!\!
			H_{2,0:3,2; \ldots ;3,2}^{0,1:1,3; \ldots ;1,3}\left(\!\!\! {\left. {\begin{array}{*{20}{c}}
						{e{A_{{0}}}\prod\limits_{\ell  = 1}^2 {\left( {\sqrt 2 {\sigma _{1,\ell }}} \right)} }\\
						\vdots \\
						{e{A_{{0}}}\prod\limits_{\ell  = 1}^2 {\left( {\sqrt 2 {\sigma _{L,\ell }}} \right)} }
				\end{array}} \!\!\right|\!\!\begin{array}{*{20}{c}}
					{\left( { - 1; - 1, \ldots , - 1} \right),\left( {0; - 1, \ldots\! , - 1} \right):{\upsilon _{FP,1}}; \ldots\! ;{\upsilon _{FP,L}}}\\
					{ - :\left( {0,1} \right),\left( { - {\gamma ^2},1} \right); \ldots \!;\left( {0,1} \right),\left( { - {\gamma ^2},1} \right)}
			\end{array}} \!\!\!\right)
	\end{align}}\noindent
		\setcounter{equation}{\value{mycount5}}
	\end{figure*}
	\addtocounter{equation}{1}
	%----------------------------------------------------------------------------------------------------------------------------
	%----------------------------------------------------------------------------------------------------------------------------

	\begin{IEEEproof}
	Please refer to Appendix \ref{ECprop}.
	\end{IEEEproof}
	\end{prop}
	\subsubsection{Ideal RF Chains}
	Exact ergodic capacity can be derived for the case of ideal RF chains.
	\begin{prop}\label{ECpropappid}
	Effective capacity of the RIS-aid system for the case of ideal RF chains can be derived as \eqref{faefae}, shown at the bottom of the next page.
	%----------------------------------------------------------------------------------------------------------------------------
	%----------------------------------------------------------------------------------------------------------------------------
	\newcounter{mycount}
	\begin{figure*}[b]
		\normalsize
		\setcounter{mycount}{\value{equation}}
		\hrulefill
		\vspace*{4pt}
	{\small \begin{align}\label{faefae}
			C{{ = }}&\sum\limits_{{j_{1,1}}, \ldots ,{j_{L,1}} = 0}^{N_T}  {\sum\limits_{{j_{1,2}}, \ldots ,{j_{L,2}} = 0}^{N_T}  {\prod\limits_{\iota  = 1}^L {\prod\limits_{\ell  = 1}^2 {\frac{{{K_{\iota ,\ell }}^{{j_{\iota ,\ell }}}{d_{\iota ,\ell }}_{{j_{\iota ,\ell }}}}}{{{j_{\iota ,\ell }}!}}\frac{{{m_{\iota ,\ell }}^{{m_{\iota ,\ell }}}}}{{\Gamma ({m_{\iota ,\ell }})}}\frac{{{\gamma ^2}}}{{\Gamma ({j_{\iota ,\ell }} + 1)}}\frac{1}{2}} } } }
			\notag\\&\times\!\!\!
			H_{2,0:3,{{2}}; \ldots ;3{{,2,2,2}}}^{0,1:1,3; \ldots ;1,3,2,1}\left(\!\! {\left. {\begin{array}{*{20}{c}}
						{\begin{array}{*{20}{c}}
								{{\eta _1}}\\
								\vdots \\
								{{\eta _L}}
						\end{array}}\\
						{\frac{1}{e}}
				\end{array}}\!\! \right|\!\!\begin{array}{*{20}{c}}
					{\left( {1; - 0.5, \ldots , - 0.5, - 1} \right),\left( {0;1, \ldots ,1,0} \right):{\upsilon _{{h_{FP}},1}}; \ldots ;{\upsilon _{{h_{FP}},L}};\left( {0,1} \right)}\\
					{ - :\left( {0,1} \right),\left( { - {\gamma ^2},1} \right); \ldots ;\left( {0,1} \right),\left( { - {\gamma ^2},1} \right);\left( {0,1} \right),\left( {0,1} \right);\left( {1,1} \right)}
			\end{array}} \!\!\right)
	\end{align}}\noindent
		\setcounter{equation}{\value{mycount}}
	\end{figure*}
	\addtocounter{equation}{1}
	%----------------------------------------------------------------------------------------------------------------------------
	%----------------------------------------------------------------------------------------------------------------------------
	
	where {\small ${\eta _\varsigma }{{ = }}{A_{{0}}}\sqrt {e\frac{{{{\left| {{h_L}} \right|}^2}P}}{{{N_o}}}} \prod\limits_{\ell  = 1}^{{2}} {\left( {\sqrt 2 {\sigma _{\varsigma ,\ell }}} \right)} $.}
	\begin{IEEEproof}
	Please refer to Appendix \ref{ECpropid}.
	\end{IEEEproof}
	\end{prop}
	\begin{prop}\label{ECpropappidup}
	Effective capacity of the RIS-aid system for the case of ideal RF chains are bounded by
	{\small \begin{equation}\label{caegfaebv}
	C_{I} \le {\log _2}\left( {1 + \frac{{P{{\left| {{h_L}} \right|}^2}{\mathbb E}\left[ {{{\left| {{h_{FP}}} \right|}^2}} \right]}}{N_o}} \right).
	\end{equation}}\noindent
	\begin{IEEEproof}
	With the help of Jensen's inequality \cite{krantz2012handbook}, the ergodic capacity can be upper-bounded as $C_{I} \le {\log _2}\left( {1 + {\mathbb E}\left[ {{\gamma _I}} \right]} \right)$. With the help of \eqref{gammaidcf}, we obtain \eqref{caegfaebv} to complete the proof.
	\end{IEEEproof}
	\end{prop}
	From \eqref{afegnyt6r}, \eqref{faefae}, and \eqref{caegfaebv}, as expected, we can observe that the multipath parameters have a positive effect on the ergodic capacity of RIS-aid THZ communications.
	
	\section{Numerical Results And Discussions}\label{nad}
	In this section, we present numerical results to investigate the joint impacts of misalignment, multipath fading, channel conditions, and transceivers hardware imperfections on the outage and ergodic capacity performance of the RIS-aided THz communications. Following the practical parameters in \cite{kokkoniemi2018simplified,boulogeorgos2021coverage,koenig2013wireless,nagatsuma2016advances,boulogeorgos2018terahertz}, we set the parameters as follows: $T=27$ $ {}^ \circ C $, $p=101325$ ${\rm Pa}$, $c=3\times10^8$ ${\rm m/s}$, $w_d/a=6$, $G_t=G_r=40$ ${\rm dBi}$, $\sigma_S=0.01$ ${\rm m}$, and the number of Monte Carlo iterations is $10^6$.
	%Moreover, for different set of channel conditions, Algorithm \ref{1} is used to obtain the corresponding optimal phase-shifts.
	
	\begin{figure}[t]
	\centering
	\includegraphics[width=0.45\textwidth]{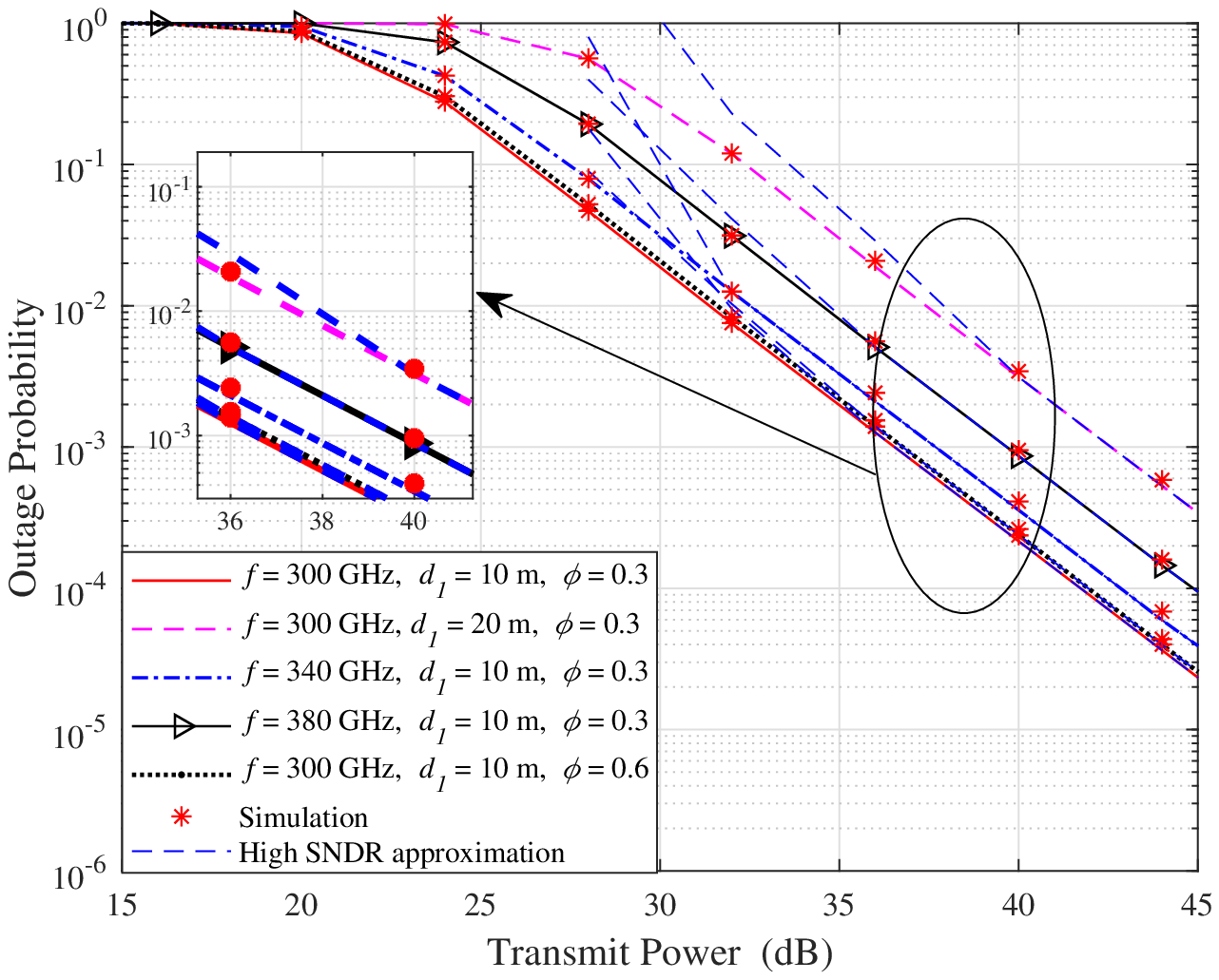}	
	\caption{Outage probability versus the transmit power with $d_2=20$ ${\rm m}$, $N_0=1$ ${\rm dBW}$, $L=40$, $m_{\ell,1}=5$, $m_{\ell,2}=7$, $K_{\ell,1}=5$, $K_{\ell,2}=6$, $\Delta_{\ell,1}=0.6$, $\Delta_{\ell,2}=0.4$, $\kappa_S=\kappa_D=0.1$, and $ {{2}}{\sigma _{\ell ,1}}\left( {1 + {K_{\ell ,1}}} \right) = {{2}}{\sigma _{\ell ,2}}\left( {1 + {K_{\ell ,2}}} \right) = 20 $ ${\rm dB}$.}
	\label{FIGOP}
\end{figure}
\begin{figure}[t]
	\centering
	\includegraphics[width=0.45\textwidth]{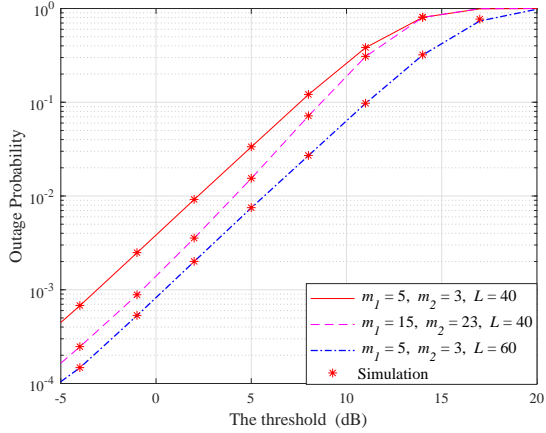}
	\caption{OP versus the threshold, $\gamma_{\rm th}$, with $P=30$ ${\rm dBW}$, $m_{\ell,1}=m_{1}$, $m_{\ell,2}=m_{2}$, $ {{2}}{\sigma _{\ell ,1}}\left( {1 + {K_{\ell ,1}}} \right) = {{2}}{\sigma _{\ell ,2}}\left( {1 + {K_{\ell ,2}}} \right) = 20$ ${\rm dB}$, $f=300$ ${\rm GHz}$, $d_1=10$ ${\rm m}$, $d_2=10$ ${\rm m}$, $N_0=1$ ${\rm dBW}$, $K_{\ell,1}=5$, $K_{\ell,2}=6$, $\Delta_{\ell,1}=0.6$, $\Delta_{\ell,2}=0.4$, and $\kappa_S=\kappa_D=0.1$.}
	\label{FIGOPth}
	\end{figure}
	
	Figure \ref{FIGOP} illustrates the outage probability performance versus the transmit power with $d_2=20$ ${\rm m}$, $N_0=0$ ${\rm dBW}$, $L=40$, $m_{\ell,1}=5$, $m_{\ell,2}=7$, $K_{\ell,1}=5$, $K_{\ell,2}=6$, $\Delta_{\ell,1}=0.6$, $\Delta_{\ell,2}=0.4$, $\kappa_S=\kappa_D=0.1$, and $ {{2}}{\sigma _{\ell ,1}}\left( {1 + {K_{\ell ,1}}} \right) = {{2}}{\sigma _{\ell ,2}}\left( {1 + {K_{\ell ,2}}} \right) = 20 $ ${\rm dB}$ $(\ell=1,\ldots,L)$. It can be observed that the approximate expressions match the exact results well in the high-SNDR regime. For a given transmission distance, we can observe that the outage probability increases as the relative humidity increases. Furthermore, the increase in the transmission distance of the signal can cause a significant increase in the OP. Comparing the impact of the propagation loss and the molecular absorption loss, we observe that the propagation loss has a more adverse impact on the system performance than that of the molecular absorption loss. The reason is that the outage probability degradation caused by increasing $d_1$ is severer than the corresponding degradation caused by increasing $\phi_1$. Moreover, the outage probability increases as the frequency increases. When the frequency increases from $340$ ${\rm GHz}$ to $380$ ${\rm GHz}$, the outage probability decreases by $58 \%$, which is faster than that when the frequency increases from $300$ ${\rm GHz}$ to $340$ ${\rm GHz}$ ($34 \%$). The reason is that a high molecular absorption exists in $380$ ${\rm GHz}$ because $380$ ${\rm GHz}$ is one of the water vapor resonance frequencies \cite{han2015multi}. Thus, it is significant for the THz communications system designers to consider the variation of the environmental conditions and the molecular absorption phenomenon.
	
	Figure \ref{FIGOPth} plots the outage probability versus the threshold, $\gamma_{\rm th}$, with $P=30$ ${\rm dBW}$, $m_{\ell,1}=m_{1}$, $m_{\ell,2}=m_{2}$, $ {{2}}{\sigma _{\ell ,1}}\left( {1 + {K_{\ell ,1}}} \right) = {{2}}{\sigma _{\ell ,2}}\left( {1 + {K_{\ell ,2}}} \right) = 20$ ${\rm dB}$, $f=300$ ${\rm GHz}$, $d_1=10$ ${\rm m}$, $d_2=20$ ${\rm m}$, $N_0=1$ ${\rm dBW}$, $K_{\ell,1}=5$, $K_{\ell,2}=6$, $\Delta_{\ell,1}=0.6$, $\Delta_{\ell,2}=0.4$, and $\kappa_S=\kappa_D=0.1$. We can observe that the outage probability increases by $72 \%$ when $\gamma_{\rm th}$ increases by $3$ ${\rm dB}$. Moreover, Fig. \ref{FIGOPth} shows the effects of different channel conditions. The outage probability decreases as shaping parameters $m_{1}$ and $m_{2}$ increase because of the better channel conditions. Furthermore, the number of RIS's reflecting elements, $L$, has a significant impact on the performance of the RIS-aided THz communications. For example, when $L$ increases from $40$ to $60$, the outage probability decreases by $97 \%$. The reason is that the RIS can effectively change the phase of reflected signals to enhance the received signal quality, enabling the desired THz wireless channel to exhibit a perfect line-of-sight.
	
	\begin{figure}[t]
	\centering
	\includegraphics[width=0.45\textwidth]{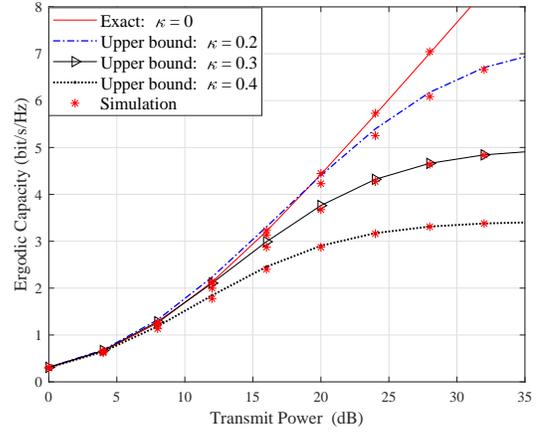}	
	\caption{Ergodic capacity versus the transmit power with $m_{\ell,1}=5$, $m_{\ell,2}=7$, $f=360$ ${\rm GHz}$, $d_1=8$ ${\rm m}$, $d_2=6$ ${\rm m}$, $N_0=0$ ${\rm dBW}$, $K_{\ell,1}=5$, $K_{\ell,2}=6$, $\Delta_{\ell,1}=0.6$, $\Delta_{\ell,2}=0.4$, $L=40$, and $ {{2}}{\sigma _{\ell ,1}}\left( {1 + {K_{\ell ,1}}} \right) = {{2}}{\sigma _{\ell ,2}}\left( {1 + {K_{\ell ,2}}} \right) = 25$ ${\rm dB}$ $(\ell=1,\ldots,L)$.}
	\label{FIGEC}
\end{figure}
\begin{figure}[t]
	\centering
	\includegraphics[width=0.45\textwidth]{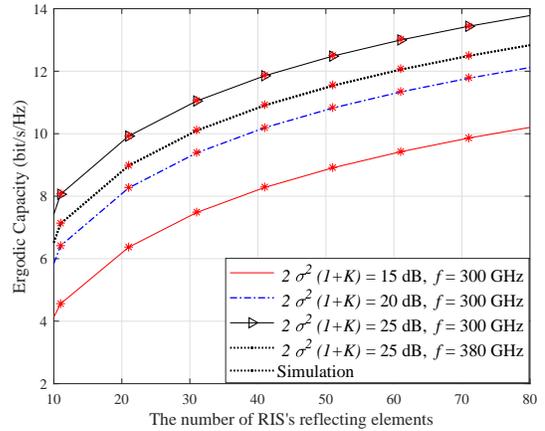}
	\caption{Ergodic capacity versus the number of RIS's reflecting elements with $\kappa_S=\kappa_D=0$, $m_{\ell,1}=5$, $m_{\ell,2}=7$, $d_1=8$ ${\rm m}$, $d_2=6$ ${\rm m}$, $N_0=1$ ${\rm dBW}$, $K_{\ell,1}=5$, $K_{\ell,2}=6$, $\Delta_{\ell,1}=0.6$, and $\Delta_{\ell,2}=0.4$.}
	\label{FIGECL}
	\end{figure}
	
	Figure \ref{FIGEC} depicts the ergodic capacity versus the transmit power with $m_{\ell,1}=5$, $m_{\ell,2}=7$, $f=360$ ${\rm GHz}$, $d_1=8$ ${\rm m}$, $d_2=6$ ${\rm m}$, $N_0=1$ ${\rm dBW}$, $K_{\ell,1}=5$, $K_{\ell,2}=6$, $\Delta_{\ell,1}=0.6$, $\Delta_{\ell,2}=0.4$, $L=40$, and $ {{2}}{\sigma _{\ell ,1}}\left( {1 + {K_{\ell ,1}}} \right) = {{2}}{\sigma _{\ell ,2}}\left( {1 + {K_{\ell ,2}}} \right) = 25$ $(\ell=1,\ldots,L)$ for different $\kappa_S=\kappa_D=\kappa$. We can observe that the ergodic capacity increases when the transmission power increases. Thus, increasing the transmission power can be regarded as a direct measure to reduce the impact of hardware imperfections. Furthermore,  Fig. \ref{FIGEC} reveals that the hardware imperfections have a small effect on the ergodic capacity in the low transmit power regime, while having a more adverse impact in the high transmit power regime. Therefore, the efficient method to increase the ergodic capacity is to decrease the hardware imperfections when the transmit power is high. In addition, we observe that the derived capacity upper bound is close to the Monte Carlo simulation results.
	
	Figure \ref{FIGECL} illustrates the ergodic capacity versus the number of RIS's reflecting elements with ${{2}}{\sigma _{\ell ,1}}\left( {1 + {K_{\ell ,1}}} \right)$ $=$ ${{2}}{\sigma _{\ell ,2}}\left( {1 + {K_{\ell ,2}}} \right)$ $=$ ${{2}}{\sigma}\left( {1 + {K}} \right) $, $\kappa_S=\kappa_D=0$, $m_{\ell,1}=5$, $m_{\ell,2}=7$, $d_1=8$ ${\rm m}$, $d_2=6$ ${\rm m}$, $N_0=1$ ${\rm dBW}$, $K_{\ell,1}=5$, $K_{\ell,2}=6$, $\Delta_{\ell,1}=0.6$, and $\Delta_{\ell,2}=0.4$ for different $f$ and ${{2}}{\sigma}\left( {1 + {K}} \right) $. From Fig. \ref{FIGECL}, we can observe that the ergodic capacity increases by $25\%$ when the number of reflecting elements increases from $40$ to $80$. This is due to the reason that more elements let RIS have superior capability in manipulating electromagnetic waves. Thus, equipping RIS with more reflecting elements can significantly improve the performance of wireless communications systems. Furthermore, the ergodic capacity decreases when the operating frequency is changed from $300$ GHz to $380$ GHz. Thus, it becomes evident that the system's performance will be significantly influenced by the selection of operating frequency, which also verifies the conclusion obtained from Fig. \ref{FIGOP}.
	
	\section{Conclusions}\label{cons}
	We investigated an RIS-aided THz communications system and presented exact expressions of the end-to-end SNR and SNDR. Based on two THz signal measurement experiments conducted in a variety of different scenarios, we proved that the FTR distribution performs a good fit for modeling the small-scale amplitude fading of THz signals. Furthermore, an SIO algorithm was proposed to find the optimal phase-shifts design at RIS under discrete phase-shift constraints. The impact of the different number of discrete phase-shift levels was investigated. Moreover, important performance metrics, including the outage probability and ergodic capacity, are derived in terms of the multivariate Fox's $H$-function. To obtain more insights, the asymptotic analysis was presented when the SNDR or the number of RIS's elements is large. The tight upper bound expressions of ergodic capacity for ideal and non-ideal RF chains are also derived. The numerical results confirmed that the propagation loss is more detrimental to the performance of the RIS-aided THz communications system than the molecular absorption loss, and the RIS can bring a significant performance gain for the proposed THz communications.
	
	\begin{appendices}
	\section{Proof of Theorem \ref{aaa111t}}\label{aaa111}
	\renewcommand{\theequation}{A-\arabic{equation}}
	\setcounter{equation}{0}
	With the help of \cite[eq. (6)]{du2020millimeter}, ${\mathbb E}\left[ {\left| {\sum\limits_{\ell  = 1}^L {{g_{1}^{\left(\ell\right)}}{g_{2}^{\left(\ell\right)}}} } \right|} \right]$ can be expressed as
	{\small \begin{equation}
	{\mathbb E}\!\left[ {\left| {\sum\limits_{\ell  = 1}^L {{g_{1}^{\left(\ell\right)}}{g_{2}^{\left(\ell\right)}}} } \right|} \right] \!\!=\! \sum\limits_{\ell  = 1}^L \prod\limits_{\ell  = 1}^2 {\frac{{m_\ell ^{{m_\ell }}}}{{\Gamma\!\left( {{m_\ell }} \right)}}} \sum\limits_{{j_\ell } = 0}^{{N_T}} {\frac{{K_\ell ^{{j_\ell }}{d_{{\ell _{{j_\ell }}}}}{{\left( {2\sigma _\ell ^2} \right)}^{\frac{1}{2}}}}}{{{j_\ell }!\Gamma\!\left( {{j_\ell } \!+ \!1} \right)}}} \Gamma\!\left(\! {\frac{3}{2} \!+\! {j_\ell }} \!\right)\!\!.
	\end{equation}}\noindent
	The $s_{\rm th}$ moment of ${\left| {{h_P}} \right|}$ is defined as $E\left[ {{{\left| {{h_P}} \right|}^s}} \right]{\rm{ = }}\int_0^{{A_o}} {{x^s}{f_{{h_p}}}(x){\rm d}x}$. With the help of \eqref{hppdfsAD}, we obtain
	{\small \begin{equation}\label{faehgWRE}
	E\left[ {{{\left| {{h_P}} \right|}^s}} \right]{{ = }}\frac{{{\gamma ^2}}}{{A_o^{{\gamma ^2}}}}\int_0^{{A_o}} {{x^{{\gamma ^2} + s - 1}}dx} =\frac{{{\gamma ^2}}}{{{\gamma ^2} + s}}{A_o}^s.
	\end{equation}}\noindent
	By substituting $s=1$ into \eqref{faehgWRE}, $E\left[ {{{\left| {{h_P}} \right|}}} \right]$ is derived. Thus, we obtain \eqref{eopt} which completes the proof.

	\section{Proof of Lemma \ref{jifenthem}}\label{jifenappen}
	\renewcommand{\theequation}{B-\arabic{equation}}
	\setcounter{equation}{0}
	\subsection{Proof of PDF}\label{appenA}
	With the help of \eqref{gammaN}, we obtain $\frac{{{1}}}{{{\gamma _N}}}{{ = }}{\kappa ^2}{{ + }}\frac{{{N_o}}}{{{{\left| {{h_L}} \right|}^2}P}}\frac{{\rm{1}}}{{{{\left| {{h_{FP}}} \right|}^2}}}$, where $ \left| {{h_{FP}}} \right| \triangleq \left| {{h_F}} \right| \times \left| {{h_P}} \right| $. Let us define that $ V \triangleq \frac{{\rm{1}}}{{{{\left| {{h_{FP}}} \right|}^2}}} $, $ W \triangleq \frac{{{N_o}}}{{{{\left| {{h_L}} \right|}^2}P}}V = \frac{{\rm{1}}}{{{\gamma _N}}} - {\kappa ^2} $. The PDF of $W$ can be respectively expressed as
	{\small \begin{align}\label{fwpdfal}
	{f_W}\left( w \right) =& \frac{{{{\left| {{h_L}} \right|}^2}P}}{{{N_o}}}{f_V}\left( {\frac{{{{\left| {{h_L}} \right|}^2}P}}{{{N_o}}}w} \right) 
	\notag \\
	= &\frac{{ - 1}}{{2{w^2}\sqrt {\frac{{{{\left| {{h_L}} \right|}^2}P}}{{{N_o}w}}} }}{f_{\left| {{h_{FP}}} \right|}}\left( {\sqrt {\frac{{{N_o}}}{{{{\left| {{h_L}} \right|}^2}Pw}}} } \right).
	\end{align}}\noindent
	By performing the change of variables $ {\gamma _N} = \frac{1}{{W + {\kappa ^2}}} $, ${f_{{\gamma _N}}}$ can be rewritten as
	{\small \begin{equation}\label{gammapdf}
	{f_{{\gamma _N}}}\!\left( x \right) \!=\! \frac{1}{{2{{\left( {1\! -\! x{\kappa ^2}} \right)}^2}\sqrt {\frac{{{{\left| {{h_L}} \right|}^2}P}}{{{N_o}\frac{{1 - x{\kappa ^2}}}{x}}}} }}{f_{\left| {{h_{FP}}} \right|}}\!\left(\! {\sqrt {\frac{{{N_o}}}{{{{\left| {{h_L}} \right|}^2}P\frac{{1 - x{\kappa ^2}}}{x}}}} } \!\right).
	\end{equation}}\noindent
	Moreover, with the help of \cite[eq. (62)]{boulogeorgos2019analytical}, we obtain the PDF of $ \left| {{h_{FP}}} \right| $ as
	{\small \begin{equation}\label{chushifs}
	{f_{\left| {{h_{FP}}} \right|}}(x) = \int_0^{{A_o}} {\frac{1}{y}} {f_{\left| {{h_F}} \right|}}\left( {\frac{x}{y}} \right){f_{\left| {{h_P}} \right|}}(y){\rm{d}}y.
	\end{equation}}\noindent
	Substituting \eqref{chushifs} into \eqref{gammapdf}, we derive the integral representation for the PDF of $ \left| {{h_{FP}}} \right| $ as \eqref{jifencaSF}.
	\subsection{Proof of CDF}
	Following similar procedures as Appendix \ref{appenA} (i.e., the proof of Lemma 1), we can derive the CDF of $V$ and $W$ as ${F_V}\left( v \right) = 1-{F_{\left| {{h_{FP}}} \right|}}\left( {\sqrt {\frac{1}{v}} } \right)$ and ${F_W}\left( w \right) = {F_V}\left( {\frac{{{{\left| {{h_L}} \right|}^2}P}}{{{N_o}}}w} \right)$. Making the change of variable $ {\gamma _N} = \frac{1}{{W + {\kappa ^2}}} $, we can derive the CDF of ${\gamma _N}$ as ${F_{{\gamma _N}}}\left( x \right) = {F_{\left| {{h_{FP}}} \right|}}\left( {\sqrt {\frac{{x{N_o}}}{{{{\left| {{h_L}} \right|}^2}P\left( {1 - x{\kappa ^2}} \right)}}} } \right)$.
	
	With the help of \cite[eq. (65)]{boulogeorgos2019analytical}, the CDF of $ \left| {{h_{FP}}} \right| $ can be expressed as
	{\small \begin{equation}\label{chushiCSFD}
	{F_{\left| {{h_{FP}}} \right|}}(x) = \int_0^{{A_o}} {{F_{\left| {{h_F}} \right|}}} \left( {\frac{x}{y}} \right){f_{\left| {{h_P}} \right|}}(y){{{\rm d}}}y.
	\end{equation}}\noindent
	Substituting \eqref{chushiCSFD} into ${F_{{\gamma _N}}}\left( x \right)$, the integral representation for the CDF of $ \left| {{h_{FP}}} \right| $ as \eqref{jifencaSF2} can be derived to complete the proof.

	\section{Proof of Theorem \ref{hfpthem}}\label{hfpappen}
	\renewcommand{\theequation}{C-\arabic{equation}}
	\setcounter{equation}{0}
	We first derive the PDF and CDF of $\left| {{h_{FP}}} \right| \buildrel \Delta \over = \left| {{h_F}} \right| \times \left| {{h_P}} \right|$.
	
	\subsection{PDF of $\left| {{h_{FP}}} \right|$}
	Using the definition of multivariate Fox's $H$-function \cite[eq. (A-1)]{mathai2009h}, we can express \cite[eq. 11]{du2020millimeter} as	
	{\small \begin{align}\label{sumpropdfzhan}
	&{f_{\left|h_{F}\right|}}(y){{ = }}\!\!\!\sum\limits_{{j_{1,1}}, \ldots ,{j_{L,1}} = 0}^{N_T} \sum\limits_{{j_{1.2}}, \ldots ,{j_{L,2}} = 0}^{{N_T}} {\prod\limits_{\iota  = 1}^L {\prod\limits_{\ell  = 1}^2 {\frac{{{K_{\iota ,\ell }}^{{j_{\iota ,\ell }}}{d_{\iota ,\ell }}_{{j_{\iota ,\ell }}}{m_{\iota ,\ell }}^{{m_{\iota ,\ell }}}}}{{{j_{\iota ,\ell }}!\Gamma\!({m_{\iota ,\ell }})\Gamma\!({j_{\iota ,\ell }} + 1)}}} } } 
	\notag\\&\times\!
	\frac{1}{y}\prod\limits_{\iota  = 1}^L \frac{1}{{2\pi i}}\int_{{{\cal L}_\ell }} \frac{{\prod\limits_{\ell  = 1}^2 {\Gamma\!\left( {1 + {j_{\iota ,\ell }} + \frac{1}{2}{\varsigma _\iota }} \right)} }}{{\Gamma\!\left( { - \sum\limits_{\iota {{ = 1}}}^L {{\varsigma _\iota }} } \right)\Gamma^{-1}\!\left( { - {\varsigma _\iota }} \right)}}
%	\notag\\&\times
	{{\left(\! {{y^{ - 1}}\prod\limits_{\ell  = 1}^2 {\left( {\sqrt 2 {\sigma _{\iota ,\ell }}} \right)} } \!\!\right)}^{{\varsigma _\iota }}} {\rm d}{\varsigma _\iota },
	\end{align}}\noindent
	where the integration path of  ${\cal L_\ell}$ $(\ell=1,2,\ldots,L)$ goes from $a_\ell -\infty j$ to $a_\ell+\infty j$ and $a_\ell  \in \mathbb{R}$. Substituting \eqref{sumpropdfzhan} and \eqref{hppdfsAD} into \eqref{chushifs}, with the help of Fubini's theorem and \cite[eq. (8.331.1)]{gradshteyn2007}, we obtain
	{\small \begin{align}\label{pdfguochengs}
	&{f_{\left| {{h_{FP}}} \right|}}(x)\! =\!\!\!\!\! \sum\limits_{{j_{1\!,\!1}}, \ldots ,{j_{L\!,\!1}} = 0}^{N_T} \! \!\!\!\ldots\! \!\!\! \sum\limits_{{j_{1\!,\!N}}, \ldots ,{j_{L\!,\!N}} = 0}^{N_T}  \prod\limits_{\iota  = 1}^L {\prod\limits_{\ell  = 1}^2 {\frac{{{K_{\iota ,\ell }}^{{j_{\iota ,\ell }}}{d_{\iota ,\ell }}_{{j_{\iota ,\ell }}}{m_{\iota ,\ell }}^{{m_{\iota ,\ell }}}}}{{{j_{\iota ,\ell }}!\Gamma\! ({m_{\iota ,\ell }})\Gamma\!({j_{\iota ,\ell }} \!+\! 1)}}} }  
	\notag\\&\times\!\!
	\frac{{{\gamma ^2}}}{{xA_o^{{\gamma ^2}}}}\prod\limits_{\iota  = 1}^L \frac{1}{{2\pi i}}\int_{{{\cal L}_\ell }} \frac{{I_A}{\prod\limits_{\ell  = 1}^2 {\Gamma\!\left(\! {1\! +\!{j_{\iota ,\ell }} \!+\! \frac{1}{2}{\varsigma _\iota }} \!\right)} }}{{\Gamma\!\left(\! { -\!\sum\limits_{\iota {\rm{ = 1}}}^L {{\varsigma _\iota }} } \!\right)\Gamma^{-1}\!\left(\! { - {\varsigma _\iota }} \!\right)}}
%	\notag\\&\times
	{{\!\left(\! {{\frac{1}{x}}\prod\limits_{\ell  = 1}^2 {\left( {\sqrt 2 {\sigma _{\iota ,\ell }}} \right)} }\! \!\right)}^{{\varsigma _\iota }}}\! \!{\rm d}{\varsigma _\iota },
	\end{align}}\noindent
	where
	{\small \begin{equation}\label{iazaisd}
	{I_A}{{ = \!\!}}\int_0^{{A_{\rm{0}}}} \!\!{{y^{{\varsigma _\iota } + {\gamma ^2} - 1}}{\rm{d}}y} {\rm{ = }}\frac{{\rm{1}}}{{{\varsigma _\iota } \!+ \!{\gamma ^2}}}{A_{\rm{0}}}^{{\varsigma _\iota } + {\gamma ^2}}=\frac{{\Gamma\!\left( {{\varsigma _\iota } \!+\! {\gamma ^2}} \right)}}{{\Gamma\!\left( {{\varsigma _\iota }\! + \!{\gamma ^2}\! +\! 1} \right)}}{A_{\rm{0}}}^{{\varsigma _\iota } \!+ \!{\gamma ^2}}.
	\end{equation}}\noindent
	Substituting ${I_A}$ into \eqref{pdfguochengs} and using \cite[eq. (A-1)]{mathai2009h}, we can obtain ${f_{\left| {{h_{FP}}} \right|}}(x)$.
	
	\subsection{CDF of $\left| {{h_{FP}}} \right|$}
	Substituting \cite[eq. 12]{du2020millimeter} and \eqref{hppdfsAD} into \eqref{chushiCSFD}, we can express the CDF of ${\left| {{h_{FP}}} \right|}$ as		
	{\small \begin{align}\label{cgccgsaol}
	&{F_{\left| {{h_{FP}}} \right|}}(x)\! = \!\!\!\!\!  \sum\limits_{{j_{1\!,\!1}}, \ldots ,{j_{L\!,\!1}} = 0}^{N_T} \!\!\!\!\ldots \!\!\!\! \sum\limits_{{j_{1\!,\!N}}, \ldots ,{j_{L\!,\!N}} = 0}^{N_T}  \prod\limits_{\iota  = 1}^L {\prod\limits_{\ell  = 1}^2 {\frac{{{K_{\iota ,\ell }}^{{j_{\iota ,\ell }}}{d_{\iota ,\ell }}_{{j_{\iota ,\ell }}}{m_{\iota ,\ell }}^{{m_{\iota ,\ell }}}}}{{{j_{\iota ,\ell }}!\Gamma\! ({m_{\iota ,\ell }})\Gamma\!({j_{\iota ,\ell }} \!+ \!1)}}} }
	\notag\\&\times	\!\!
	\frac{{{\gamma ^2}}}{{A_o^{{\gamma ^2}}}}
	\prod\limits_{\iota  = 1}^L \frac{1}{{2\pi i}}\int_{{{\cal L}_\iota }} \!\!\frac{{I_A}{\prod\limits_{\ell  = 1}^2 {\Gamma \!\left( {1\! +\! {j_{\iota ,\ell }} \!+ \!\frac{1}{2}{\varsigma _\iota }} \right)} }}{{\Gamma\!\left(\! {1 \!-\! \sum\limits_{\iota {\rm{ = 1}}}^L {{\varsigma _\iota }} } \!\right)\Gamma^{-1}\!\left(\! { - {\varsigma _\iota }} \!\right)}}\!
	{{\left(\!\! {\frac{{\rm{1}}}{x}\prod\limits_{\ell  = 1}^2 {\left( {\sqrt 2 {\sigma _{\iota ,\ell }}} \right)} } \!\!\right)}^{{\varsigma _\iota }}}{\rm d}{\varsigma _\iota },
	\end{align}}\noindent
	where ${I_A}$ has been solved in \eqref{iazaisd}. Substituting  ${I_A}$ into \eqref{cgccgsaol}, the CDF of ${\left| {{h_{FP}}} \right|}$ can be expressed in terms of multivariate Fox's $H$-function. Therefore, with the help of \eqref{jifencaSF}, \eqref{jifencaSF2}, \eqref{pdfguochengs}, and ${F_{\left| {{h_{FP}}} \right|}}(x)$, the PDF and CDF of ${{\gamma _N}}$ can be derived as \eqref{GAMMANPDF} and \eqref{GAMMANCDF}, respectively.

	\section{Proof of Proposition \ref{newprop}}\label{newproof}
	\renewcommand{\theequation}{D-\arabic{equation}}
	\setcounter{equation}{0}
	To derive the approximate PDF of $\gamma_N$, we need first to get the PDF of $\left\| {{{\bf{h}}_F}} \right\|$, where $ \left\| {{{\bf{h}}_F}} \right\| = \sum\limits_{\ell  = 1}^L {\left| {g_2^{\left( \ell  \right)}} \right|\left| {g_1^{\left( \ell  \right)}} \right|}  $. With the help of a 5FM based method proposed in \cite{el2018accurate} and the definition of Meijer's $G$ function \cite[eq. (9.30)]{gradshteyn2007}, we can obtain the approximate PDF of $\left\| {{{\bf{h}}_F}} \right\|$ as
	{\small \begin{equation}\label{cahjk}
	{F_{\left| {{h_F}} \right|}} = \frac{{{a_1}}}{{2\pi i}}\int_{\cal L} {\frac{{x\Gamma \left( {{a_4} - s} \right)\Gamma \left( {{a_5} - s} \right)\Gamma \left( {s + 1} \right)}}{{\Gamma \left( {{a_3} - s} \right)\Gamma \left( {s + 2} \right)}}} {\left( {\frac{x}{{{a_2}}}} \right)^s}{\rm d}s.
	\end{equation}}\noindent
	Substituting \eqref{cahjk} and \eqref{hppdfsAD} into \eqref{jifencaSF2}, we have
	{\small \begin{align}\label{ageg34}
	{F_{{\gamma _N}}}\left( x \right)& = {\gamma ^2}A_0^{ - {\gamma ^2}}\frac{{{a_1}}}{{2\pi i}}\int_{\cal L} {\frac{{\Gamma \left( {{a_4} - s} \right)\Gamma \left( {{a_5} - s} \right)\Gamma \left( {s + 1} \right)}}{{\Gamma \left( {{a_3} - s} \right)\Gamma \left( {s + 2} \right)}} }
	\notag\\&\times	
	\Upsilon{\left( {\frac{1}{{{a_2}}}\Upsilon } \right)^s}{I_D}{\rm d}s,
	\end{align}}\noindent
	where $I_D$ can be solved with the help of \cite[eq. (8.331.1)]{gradshteyn2007} as
	{\small \begin{equation}
	{I_D}{\rm{ = }}\!\int_0^{{A_o}} \!{{y^{ - s - 2 + {\gamma ^2}}}{\rm{d}}y}  =\! \frac{{{A_o}^{ - 1 + {\gamma ^2} - s}}}{{ - 1 \!+ \!{\gamma ^2} \!-\! s}} = \frac{{{A_o}^{ - 1 + {\gamma ^2} - s}\Gamma\!\left( {{\gamma ^2} - s - 1} \right)}}{{\Gamma\!\left( {{\gamma ^2} - s} \right)}}.
	\end{equation}}\noindent
	Substituting $I_D$ into \eqref{ageg34} and using \cite[eq. (9.30)]{gradshteyn2007}, we can obtain \eqref{qwerlk} to complete the proof.

	\section{Proof of Proposition \ref{ECpropapp}}\label{ECprop}
	\renewcommand{\theequation}{E-\arabic{equation}}
	\setcounter{equation}{0}
	With the help of Jensen's inequality \cite{krantz2012handbook}, the upper-bound of the ergodic capacity can be expressed as $C_{N} \le {\log _2}\left( {1 + {\mathbb E}\left[ {{\gamma _N}} \right]} \right)$, where
	{\small 	\begin{equation}\label{agfebntry}
	{\mathbb E}\left[ {{\gamma _N}} \right]\!{{ = }}{\mathbb E}\left[\! {\frac{{{{\left| {{h_{FP}}} \right|}^2}{{\left| {{h_L}} \right|}^2}P}}{{{\kappa ^2}{{\left| {{h_{FP}}} \right|}^2}{{\left| {{h_L}} \right|}^2}P + {N_o}}}} \!\right]\!{\rm{ = }}\frac{{P{{\left| {{h_L}} \right|}^2}{\mathbb E}\left[ {{{\left| {{h_{FP}}} \right|}^2}} \right]}}{{P{\kappa ^2}{{\left| {{h_L}} \right|}^2}{\mathbb E}\left[ {{{\left| {{h_{FP}}} \right|}^2}} \right] + {N_o}}}.
	\end{equation}}\noindent
	The $s_{\rm th}$ moment of ${\left| {{h_{FP}}} \right|}$ is defined as ${\mathbb E}\left[ {{{\left| {{h_{FP}}} \right|}^s}} \right]{{ = }}\int_0^{\infty}  {{x^s}{f_{{h_{FP}}}}(x){\rm d}x}$. With the help of \eqref{pdfguochengs}, ${\mathbb E}\left[ {{{\left| {{h_{FP}}} \right|}^s}} \right]$ can be expressed as \eqref{exphfpsd}, shown at the bottom of the next page, where {\small ${I_C} = \int_0^{\infty} {{x^{s - \sum\limits_{\iota  = 1}^L {{\varsigma _\iota }}  - 1}}{\rm d}x}$}.

		%----------------------------------------------------------------------------------------------------------------------------
	%----------------------------------------------------------------------------------------------------------------------------
	\newcounter{mycoun}
	\begin{figure*}[b]
		\normalsize
		\setcounter{mycoun}{\value{equation}}
		\hrulefill
		\vspace*{4pt}
	{\small \begin{align}\label{exphfpsd}
			{\mathbb E}\left[ {{{\left| {{h_{FP}}} \right|}^s}} \right]{{ = }}&\sum\limits_{{j_{1,1}}, \ldots ,{j_{L,1}} = 0}^{N_T}   \ldots  \sum\limits_{{j_{1,N}}, \ldots ,{j_{L,N}} = 0}^{N_T}  {\prod\limits_{\iota  = 1}^L {\prod\limits_{\ell  = 1}^2 {\frac{{{K_{\iota ,\ell }}^{{j_{\iota ,\ell }}}{d_{\iota ,\ell }}_{{j_{\iota ,\ell }}}}}{{{j_{\iota ,\ell }}!}}\frac{{{m_{\iota ,\ell }}^{{m_{\iota ,\ell }}}}}{{\Gamma ({m_{\iota ,\ell }})}}\frac{{{\gamma ^2}}}{{\Gamma ({j_{\iota ,\ell }} + 1)}}} } }
			\notag\\&\times
			\prod\limits_{\iota  = 1}^L {\frac{1}{{2\pi i}}\int_{{{\cal L}_\ell }} {\frac{{\prod\limits_{\ell  = 1}^2 {\Gamma \left( {1 + {j_{\iota ,\ell }} + \frac{1}{2}{\varsigma _\iota }} \right)} \Gamma \left( { - {\varsigma _\iota }} \right)\Gamma \left( {{\varsigma _\iota } + {\gamma ^2}} \right)}}{{\Gamma \left({ -\sum\limits_{\iota {\rm{ = 1}}}^L {{\varsigma _\iota }} } \right)\Gamma \left( {{\varsigma_\iota } + {\gamma ^2} + 1} \right)}}{{\left( {{A_{\rm{0}}}\prod\limits_{\ell  = 1}^2 {\left( {\sqrt 2 {\sigma _{\iota ,\ell }}} \right)} } \right)}^{{\varsigma_\iota }}}{I_C}} {\rm d}{\varsigma_\iota }}
	\end{align}}\noindent
		\setcounter{equation}{\value{mycoun}}
	\end{figure*}
	\addtocounter{equation}{1}
	%----------------------------------------------------------------------------------------------------------------------------
	%----------------------------------------------------------------------------------------------------------------------------

	Let ${\cal L}\left\{ {p\left( t \right)} \right\} = P\left( x \right)$. With the help of the property of Laplace transform, it follows that ${\cal L}\left\{ {\int_0^t {p\left( z \right)dz} } \right\} = \frac{{P\left( x \right)}}{x}$. According to the final value theorem, we have
	{\small \begin{equation}\label{finaltheorem}
	\mathop {\lim }\limits_{t \to {N_T} } \left( {\int_0^t {p\left( z \right)dz} } \right) = e\frac{{P\left( e \right)}}{e} = P\left( e\right),
	\end{equation}}\noindent
	where $e$ is a number close to zero (e.g., $e=10^{-6}$). Thus, ${I_C}$ can be solved as {\small ${I_C}  = {\left( {\frac{1}{e}} \right)^{s - \sum\limits_{\iota  = 1}^L {{\varsigma _\iota }} }}\Gamma \left( {s - \sum\limits_{\iota  = 1}^L {{\varsigma _\iota }} } \right)$}. Substituting ${I_C}$ into \eqref{exphfpsd} and using \cite[eq. (A-1)]{mathai2009h}, we can rewrite {\small ${\mathbb E}\left[ {{{\left| {{h_{FP}}} \right|}^s}} \right]$} in terms of multivariate Fox's $H$-function. With the help of \eqref{agfebntry}, letting $s=2$, we obtain \eqref{afegnyt6r} to complete the proof.
	
	\section{Proof of Proposition \ref{ECpropappid}}\label{ECpropid}
	\renewcommand{\theequation}{F-\arabic{equation}}
	\setcounter{equation}{0}
	For the case of ideal RF chains, the ergodic capacity can be expressed as ${C_I} = \int_0^{\infty}  {{{\log }_2}\left( {1 + x} \right)} {f_{\left| {{\gamma _I}} \right|}}\left( x \right){\rm d}x$. Substituting \eqref{qy2uh4i312} into ${C_I}$ and using \cite[eq. (01.04.07.0002.01)]{web}, we obtain
	{\small \begin{align}\label{afvebvaes}
	C_{I}{{ = }}&\frac{1}{2}\sum\limits_{{j_{1,1}}, \ldots ,{j_{L,1}} = 0}^{N_T}  {\sum\limits_{{j_{1,2}}, \ldots ,{j_{L,2}} = 0}^{N_T}  {\prod\limits_{\iota  = 1}^L \prod\limits_{\ell  = 1}^2 {\frac{{{K_{\iota ,\ell }}^{{j_{\iota ,\ell }}}{d_{\iota ,\ell }}_{{j_{\iota ,\ell }}}{m_{\iota ,\ell }}^{{m_{\iota ,\ell }}}{\gamma ^2}}}{{{j_{\iota ,\ell }}!\Gamma ({m_{\iota ,\ell }})\Gamma ({j_{\iota ,\ell }} + 1)}}}  } }
	\notag\\&\times
	\prod\limits_{\iota  = 1}^L \frac{1}{{2\pi i}}\int_{{{\cal L}_\ell }} \frac{{\prod\limits_{\ell  = 1}^2 {\Gamma \left( {1 + {j_{\iota ,\ell }} + \frac{1}{2}{\varsigma _\iota }} \right)} \Gamma \left( { - {\varsigma _\iota }} \right)\Gamma \left( {{\varsigma _\iota } + {\gamma ^2}} \right)}}{{\Gamma \left( { - \sum\limits_{\iota {\rm{ = 1}}}^L {{\varsigma _\iota }} } \right)\Gamma \left( {{\varsigma _\iota } + {\gamma ^2} + 1} \right)}}
	\notag\\&\times
	{{\left( {{A_{{0}}}\sqrt {\frac{{{{\left| {{h_L}} \right|}^2}P}}{{{N_o}}}} \prod\limits_{\ell  = 1}^{{2}} {\left( {\sqrt 2 {\sigma _{\iota ,\ell }}} \right)} } \right)}^{{\varsigma _\iota }}}{I_{D_1}} {\rm d}{\varsigma _\iota },
	\end{align}}\noindent
	where 
{\small \begin{align}
			{I_{D_1}} = &\int_0^{\infty} {{{\log }_2}\left( {1 + x} \right)} {x^{ - 1 - \frac{1}{2}\sum\limits_{\iota {{ = 1}}}^L {{\varsigma _\iota }} }}dx 
			\notag \\ &
			{{ = }}\frac{1}{{2\pi i}}\int_{{\cal L}_{L+1}} {\frac{{\Gamma \left( {{\varsigma _{L + 1}} + 1} \right){\Gamma ^2}\left( { - {\varsigma _{L + 1}}} \right)}}{{\Gamma \left( {1 - {\varsigma _{L + 1}}} \right)}}{I_{{D_{{2}}}}}} {\rm d}{\varsigma _{L + 1}},
	\end{align}}\noindent
and $I_{D_2}$ can be solved by following similar procedures as \eqref{finaltheorem} as 
{\small \begin{align}
		{I_{D_2}} \!= & \int_0^{\infty} {{x^{ - 1 - \frac{1}{2}\sum\limits_{\iota {{ = 1}}}^L {{\varsigma _\iota }}  - {\varsigma _{L + 1}}}}} {\rm d}x 
		\notag\\&\times
		= {\left(\! {\frac{1}{e}} \!\right)^{ - \frac{1}{2}\sum\limits_{\iota {{ = 1}}}^L {{\varsigma _\iota }}  - {\varsigma _{L + 1}}}}\Gamma\!\left(\! { - \frac{1}{2}\sum\limits_{\iota {{ = 1}}}^L {{\varsigma _\iota }} \! - {\varsigma _{L + 1}}} \right).
		\end{align}}
Substituting ${I_{D_2}}$ and ${I_{D_1}}$ into \eqref{afvebvaes}, we obtain 
	{\small \begin{align}\label{asdfvcdv}
	{C_I}{{ = }}&\sum\limits_{{j_{1,1}}, \ldots ,{j_{L,1}} = 0}^{N_T}  \sum\limits_{{j_{1,2}}, \ldots ,{j_{L,2}} = 0}^{{N_T}} {\prod\limits_{\iota  = 1}^L {\prod\limits_{\ell  = 1}^2 {\frac{{{K_{\iota ,\ell }}^{{j_{\iota ,\ell }}}{d_{\iota ,\ell }}_{{j_{\iota ,\ell }}}{m_{\iota ,\ell }}^{{m_{\iota ,\ell }}}{\gamma ^2}}}{{2{j_{\iota ,\ell }}!\Gamma ({m_{\iota ,\ell }})\Gamma ({j_{\iota ,\ell }} + 1)}}} } } 
	\notag\\&\times
	\prod\limits_{\iota  = 1}^L \frac{1}{{2\pi i}}\int_{{{\cal L}_\ell }} \frac{{\prod\limits_{\ell  = 1}^2 {\Gamma \left( {1 + {j_{\iota ,\ell }} + \frac{1}{2}{\varsigma _\iota }} \right)} \Gamma \left( { - {\varsigma _\iota }} \right)\Gamma \left( {{\varsigma _\iota } + {\gamma ^2}} \right)}}{{\Gamma \left( { - \sum\limits_{\iota {{ = 1}}}^L {{\varsigma _\iota }} } \right)\Gamma \left( {{\varsigma _\iota } + {\gamma ^2} + 1} \right)}}
	\notag\\&\times
	{{\left( {{A_{\rm{0}}}\sqrt {e\frac{{{{\left| {{h_L}} \right|}^2}P}}{{{N_o}}}} \prod\limits_{\ell  = 1}^{\rm{2}} {\left( {\sqrt 2 {\sigma _{\iota ,\ell }}} \right)} } \right)}^{{\varsigma _\iota }}} \frac{1}{{2\pi i}}
	\notag\\&\times
\int_{{L_{L + 1}}} {\frac{{{\Gamma ^2}\left( { - {\varsigma _{L + 1}}} \right)\Gamma \left( { - \frac{1}{2}\sum\limits_{\iota  = 1}^L {{\varsigma _\iota }}  - {\varsigma _{L + 1}}} \right)}}{{{\Gamma ^{ - 1}}\left( {{\varsigma _{L + 1}} + 1} \right)\Gamma \left( {1 - {\varsigma _{L + 1}}} \right){e^{{\varsigma _{L + 1}}}}}}} {\rm{d}}{\varsigma _{L + 1}}{\rm{d}}{\varsigma _\iota }.
	\end{align}}\noindent
	With the help of \cite[eq. (A-1)]{mathai2009h}, we can rewrite \eqref{asdfvcdv} as \eqref{faefae} that completes the proof.

	\end{appendices}
	
	\bibliographystyle{IEEEtran}
	\bibliography{IEEEabrv,Ref}

	\end{document}